\documentclass[acmsmall,nonacm,screen,natbib=false]{acmart}

\usepackage{comment}
\includecomment{exclude}
\excludecomment{exclude}

\usepackage[utf8]{inputenc}
\usepackage[T1]{fontenc}

\usepackage{url}
\usepackage{hyperref}

\RequirePackage[
  datamodel=acmdatamodel,
  url=false,
  maxcitenames=2,
  maxbibnames=9,
  giveninits,
  ]{biblatex}
\AtBeginBibliography{\footnotesize}

\usepackage[raster]{tcolorbox}
\usepackage[noline,noend]{algorithm2e}
\usepackage{graphicx}

\usepackage{multirow}
\RequirePackage{multicol}
\usepackage{booktabs}

\usepackage{environ}

\RequirePackage{amsmath, amsthm, mathtools}
\input{mathdefs}

\definecolor{light-gray}{gray}{0.95}
\newenvironment{boxes}[1]
{\begin{tcbraster}[raster columns=#1, raster equal height, size=small,
  colframe=black!3!white,
  colback=white,
  coltitle=black
  ]}
{\end{tcbraster}}

\newcommand\topic[1]{\paragraph{#1}}
\newcommand\sep{\vfill}

\newcommand\concurrent{\mathrel{{|}{|}}}

\auxfun{prepare}
\auxfun{effect}
\auxfun{eval}
\auxfun{apply}
\auxfun{merge}
\auxfun{inc}
\auxfun{dec}
\auxfun{add}
\auxfun{rmv}
\auxfun{ins}
\auxfun{del}
\auxfun{cbcast}
\auxfun{tcbcast}
\auxfun{tcdeliver}
\auxfun{tcstable}
\auxfun{stable}
\auxfun{stabilize}
\auxfun{delivered}
\auxfun{origin}
\auxfun{obsolete}
\auxfun{operation}
\newcommand\polog{PO-Log\xspace}

\newenvironment{algo}[1]{
  \begin{algorithm}[H]
  \DontPrintSemicolon
  \SetKwBlock{types}{types:}{}
  \SetKwBlock{gstate}{global replica state:}{}
  \SetKwBlock{cstate}{CRDT state:}{}
  \SetKwBlock{query}{query}{}
  \SetKwBlock{update}{update}{}
  \SetKwBlock{parameters}{parameters:}{}
  \SetKwBlock{dstate}{durable state:}{}
  \SetKwBlock{vstate}{volatile state:}{}
  \SetKwBlock{periodically}{periodically}{}
  \SetKwBlock{on}{on}{}
  \SetKw{return}{return}
  \SetKw{kwlet}{let}
  \SetKw{kwin}{in}
  \SetKwBlock{prepare}{prepare}{}
  \SetKwBlock{effect}{effect}{}
  \ifnum#1<2\relax\else\begin{multicols}{#1}\fi
  \newcommand\algocolumns{#1}
}{
  \ifnum\algocolumns<2\relax\else\end{multicols}\fi
  \end{algorithm}
}

\begin{document}

\title{Approaches to Conflict-free Replicated Data Types}
\author{Paulo Sérgio Almeida}
\email{psa@di.uminho.pt}
\orcid{0000-0001-7000-0485}
\affiliation{%
  \institution{INESC TEC \& University of Minho}
  \country{Portugal}
}

\begin{abstract}
  Conflict-free Replicated Data Types (CRDTs) allow optimistic replication in a
  principled way. Different replicas can proceed independently, being
  available even under network partitions, and always converging deterministically:
  replicas that have received the same updates will have equivalent state,
  even if received in different orders. After a historical tour of the
  evolution from sequential data types to CRDTs, we present in detail the two
  main approaches to CRDTs, operation-based and state-based, including two important
  variations, the pure operation-based and the delta-state based.
  Intended for prospective CRDT researchers and designers, this paper provides solid
  coverage of the essential concepts, clarifying some misconceptions which
  frequently occur, but also presents some novel insights gained from
  considerable experience in designing both specific CRDTs and approaches to CRDTs.
\end{abstract}



\maketitle

\section{Introduction}

Classic distributed systems aim for strong consistency (e.g.,
linearizability~\cite{DBLP:journals/toplas/HerlihyW90}), through the
state-machine replication approach, proposed
by~\textcite{DBLP:journals/cacm/Lamport78}. But while they can achieve
performance (throughput), achieving strong consistency in systems with large
spatial spans comes at the cost of high response time and the loss of
availability under network partitions, as expressed by the CAP
theorem~\cite{DBLP:conf/podc/Brewer00,DBLP:journals/sigact/GilbertL02}.

The importance of always-on availability for real businesses motivated
relaxing consistency. A seminal work, which popularized the NoSQL movement,
was Amazon's Dynamo~\cite{DBLP:conf/sosp/DeCandiaHJKLPSVV07}, which stresses
the importance of availability: ``even the slightest outage has significant financial
consequences and impacts customer trust''. But programming NoSQL data
stores is difficult and error prone, given a low level read-write based API,
and the need for ad hoc reconciliation of concurrent updates.

Since their appearance~\cite{DBLP:conf/sss/ShapiroPBZ11}, Conflict-free
Replicated Data Types (CRDTs), soon became very popular.
The essential concept is 1) providing a higher level API, as in classic data types,
but for distributed, replicated objects, while 2) achieving availability through
relaxing consistency and allowing immediate local replica updates and queries,
with asynchronous communication to make replicas converge, 3) with data type
specific concurrency semantics and synchronization specified as built-in,
freeing programmers from writing ad hoc reconciliation code.

CRDTs are difficult to design, prone to subtle bugs, but allow the majority of
distributed application programmers to use them, from some library, with
little effort, while only a minority of expert CRDT designers need to go
through the intricate process of creating new CRDTs over time. This paper is
mostly aimed at prospective CRDT researchers or designers, but every CRDT user
gains from having some knowledge about how they work. It describes the two main
approaches to CRDTs, based on propagating operations or on propagating state,
while also presenting two variants. The most important, delta-state
CRDTs~\cite{DBLP:journals/jpdc/AlmeidaSB18}, aim to achieve, in a way, ``the
best of both worlds''. Pure operation-based
CRDTs~\cite{DBLP:conf/dais/BaqueroAS14,DBLP:journals/corr/abs-1710-04469}
are a relevant point in the design space that makes clear the role of
specifications over a partially ordered set of operations (a partially ordered
log) in the definition of the CRDT and of \emph{causal stability} in achieving
a small state, not achievable otherwise.

This paper clarifies misconceptions regarding CRDTs, which
frequently occur, in papers or presentations, and shows novel
depictions, e.g., to provide intuition about joining
causal state-based CRDTs. One common misconception is regarding
commutativity: ``CRDTs are types with commutative operations'', which is
not true. Indeed, the more significant improvement over classic optimistic
replication, like Lazy Replication~\cite{DBLP:journals/tocs/LadinLSG92} is
the support for data types with non-commutative operations. We clarify the
role of commutativity,  by presenting a better (than the usual) diagram showing
the execution model of operation-based CRDTs.
Another example is regarding monotonicity in state-based CRDTs: ``mutators
must be monotonic functions'', when in fact they must be \emph{inflations}, to
result in the monotonic evolution of state.
The paper discusses the role of commutativity, idempotence and
inflations in the ability to reuse sequential data types for
both operation- and state-based CRDT designs.
The classic requirement of \emph{prepare} in operation-based CRDTs to be
side-effect free is also addressed. We point out that if we distinguish the
abstract state used in queries from the full CRDT concrete state, then the
requirement can be relaxed, leading to better designs, and present a novel
\emph{observed-remove set} which is better than any prior design.

After a historical tour showing the evolution from sequential data types
to CRDTs, the subsequent sections present: operation-based CRDTs, pure
operation-based CRDTs, state-based CRDTs, and delta-state based CRDTs. This is
followed by a comparison of the approaches, a discussion of identity
management towards scalability, and a presentation of practical applications.
Along the paper we use classic examples (counters, registers, sets), which are
relatively simple to explain and understand, while having enough subtlety to
allow comparing approaches, avoiding more complex data types, such as lists
(e.g., Treedoc~\cite{DBLP:conf/icdcs/PreguicaMSL09},
RGA~\cite{DBLP:journals/jpdc/RohJKL11}), which need considerably more involved
algorithms. Collaborative Editing, and the List data type are discussed in
Section~\ref{sec:practical-applications}, Practical Applications.

\section{From sequential data types to CRDTs}

\subsection{From sequential to concurrent data abstractions}

\topic{Data abstractions}

Abstraction is essential to tame complexity and scale.
Two main types are functional and data abstraction. Functional abstractions
(functions and procedures) were
introduced first, roughly at the same time (1958) in Fortran, Lisp and Algol.
Data abstractions involved a longer evolution over time, with the two main
variants being abstract data types and objects.

The ingredients to obtain data types are normally thought of as: procedures; the
concept of record, introduced in the AED-1
language~\cite{DBLP:conf/dac/RossF64} (then named \emph{plexes}) and adopted
for Algol by~\textcite{DBLP:journals/cacm/WirthH66}; the
combination of procedures and records. What actually
happened~\cite{DBLP:journals/sigplan/NygaardD78} was the
generalization of Algol blocks to lifetime not restricted to stack allocation
in Simula~\cite{DBLP:journals/cacm/DahlN66} processes (and later classes).
The final ingredient was hiding the representation from client code, in
CLU~\cite{DBLP:journals/sigplan/LiskovZ74}, leading to abstract data types
(ADTs). (Simula had some information hiding capability, through inner blocks,
which was not adequate.)

\topic{Sequential data types}

In imperative languages, sequential data types became the most common
abstraction in libraries. The availability of types such as Set and Map,
serving as ``Swiss Army knifes'' in solving many problems diminished the number of
times programmers had to ``reinvent the wheel'' and end up with slow and
buggy implementations. The research effort in efficient imperative
implementations of such data types has been immensely useful for the ``real
world''.

For sequential data types proving correctness is relatively easy (when no
aliasing is involved), made possible with the introduction of axiomatic
reasoning by~\textcite{DBLP:journals/cacm/Hoare69}, involving the notions of
preconditions, postconditions and invariants, and its adaptation to data
types~\cite{DBLP:journals/acta/Hoare72}. The pervading occurrence of aliasing
poses a problem; languages like Rust, aim to avoiding mutable state
sharing.

\begin{exclude}
While easier to reason about, they are harder to implement efficiently in
functional languages, due to the absence of destructive updates and the need
for persistence. Amortization of cost can be achieved with lazy
evaluation~\cite{DBLP:books/daglib/0097014}, but still not matching what is
possible in imperative implementations, and causing performance
unpredictability.
\end{exclude}

\topic{Concurrent data types}

It was only natural to extend sequential data types to shared memory
concurrency, to obtain objects usable by concurrent threads. Inspired by
Simula, \textcite{10.5555/540365} introduced
monitors, with the construct of \emph{shared classes},
and~\textcite{DBLP:journals/cacm/Hoare74} introduced a slight variant.
Monitors enforce mutual-exclusion during operation execution, allowing the
concept of \emph{atomic objects}, which are easy to reason about using the
same concepts of pre-/post-conditions and invariants, as no concurrency occurs
during operation execution.
Monitors also allow blocking mid-operation, via the await primitive (Hansen),
or condition variables (Hoare). This is something useful and
essential for inter-process synchronization, to achieve cooperation through
shared abstractions, the most well known being the \emph{bounded-buffer}.
Unfortunately, it complicates reasoning. Blocking abstractions, common in
shared-memory concurrency, are less common in distributed systems (an example
being a distributed lock), and do not suit CRDTs which aim to be always
available locally.

\topic{Lock-free and wait-free data structures}

Towards achieving performance and fault tolerance in shared-memory
multiprocessors, lock-free data structures were proposed. These do no use
locks but resort directly to low level atomic operations, such as
\emph{compare-and-swap}, provided by hardware. This allows several data type
operations in progress concurrently, not in mutual exclusion (as in
monitors). Moreover, if wait-free~\cite{DBLP:journals/toplas/Herlihy91}, it
guarantees that an operation completes in a finite number of steps, regardless
of what other processes do. Their emphasis is on multiprocessors, not
distributed systems, not being of further concern here, except for one
question they arise: how to express correctness criteria when several actions
are happening concurrently. This issue is relevant for distributed systems.

\subsection{Strongly consistent replication}

\topic{Linearizability}

Linearizability~\cite{DBLP:journals/toplas/HerlihyW90} is the more widely used
correctness criteria when aiming for implementations that, even allowing
concurrent execution of operations, mimic the behavior exposed by a sequential
data type. An execution is linearizable, roughly, if 1) it is equivalent to some
sequential execution; 2) it respects ``real time'' for non-overlapping operations.

\begin{exclude}
\begin{quote}
  Linearizability provides the illusion that each \textbf{operation}
  applied by concurrent processes takes effect \textbf{instantaneously} at
  some point \textbf{between} its \textbf{invocation} and its
  \textbf{response} (\textcite{DBLP:journals/toplas/HerlihyW90})
\end{quote}
\end{exclude}

\begin{figure}
\includegraphics[scale=0.8]{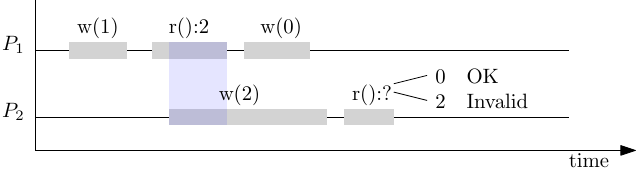}
  \caption{A register object with read and write operations. To be
  linearizable, last read must return 0.}
  \label{fig:linearizability}
\end{figure}

Consider a register object (Figure~\ref{fig:linearizability})
being accessed by two processes. The read from $P_2$ must return 0 if
the register provides linearizability, even though w(2) was the last operation
to return, before the read. The reason is that because the read from $P_1$
returned 2, the write from $P_2$ must have taken effect before it, in the
interval shaded in blue, being ordered before r():2 and w(0) from $P_1$.

\topic{Replicated state machines}

Obtaining distributed implementations of data types that comply with
linearizability can be done through the state-machine replication approach.
This was proposed by Lamport~\cite{DBLP:journals/cacm/Lamport78} and consists
essentially of:
  \begin{enumerate}
    \item Replicate state on several nodes;
    \item Same deterministic state machine at each node, $S' = F(S, I)$, where
      the next state is a function of the current one and the input;
    \item Agree on a global total order of inputs from all nodes;
    \item Apply totally ordered inputs at each node.
  \end{enumerate}

The result is that the replicated system behaves as if it were a single
machine. The most difficult part, specially to be fault tolerant, is the
agreement~\cite{DBLP:journals/jacm/PeaseSL80}. The extreme case of tolerating
byzantine faults is applied in the now popular blockchains, many of which use
some variant of the PBFT algorithm~\cite{DBLP:conf/osdi/CastroL99}.
Using a sequential data type for the state machine results in a replicated
data type. One could think then that the problem of obtaining distributed
implementations of data types is solved and there is nothing more that needs
to be done.

\subsection{The CAP theorem and consistency models}

\topic{The CAP theorem}
In a keynote at the PODC 2000 conference, \textcite{DBLP:conf/podc/Brewer00}
stated a conjecture that in a distributed system we can
only achieve, simultaneously, at most two of the three guarantees:
strong Consistency, Availability, Partition tolerance.
This conjecture was proved~\cite{DBLP:journals/sigact/GilbertL02}, more
concretely, interpreting strong consistency as linearizability, and became known
as the CAP theorem. This is commonly expressed as a trilemma, in which we can
only pick two out of the three properties. 
But because network partitions may always occur, and cannot be avoided, we can
design either AP or CP systems: achieve either availability or strong
consistency (linearizability).

\topic{CP vs AP}

The CAP theorem implies an important design choice for distributed systems.
Whether to achieve linearizability (CP) or availability (AP).
But even when there are no network partitions, the
design choice has also implication on response times. As a rough summary, CP
systems:
    \begin{itemize}
      \item aim for linearizability;
      \item may become unavailable under network partitions;
      \item have high response times in wide area.
    \end{itemize}
In AP systems:
    \begin{itemize}
      \item operations can remain available, even when there are partitions;
      \item response times can be low even in wide area.
    \end{itemize}

Forgoing linearizability, an important question is: what consistency model to aim for?

\topic{Consistency models}

The definition of consistency models has been an important and complex
research topic going over many decades. It typically involves tradeoffs
regarding consequences to relevant actors (e.g., hardware, compiler,
programmers), in matters such as efficient use of ``the machine'', ease of
implementability, ease of reasoning, or useful guarantees.
\textcite{DBLP:journals/csur/ViottiV16} present over 40 models.
Of these, causal consistency~\cite{DBLP:journals/dc/AhamadNBKH95} is of
special importance, being (broadly) the strongest achievable while not losing
availability~\cite{MADahlin2011}.

\begin{exclude}
Figure~\ref{fig:consistency-models} presents some important consistency
models/guarantees for distributed systems, the strongest on top.
\textcite{DBLP:journals/csur/ViottiV16} present over 40 models, not even
including transactional models. The horizontal line separates what is
achievable in AP systems and in CP systems. Of these, causal
consistency~\cite{DBLP:journals/dc/AhamadNBKH95} is of special importance,
being (broadly) the strongest achievable while not losing
availability~\cite{MADahlin2011}.

\begin{figure}
  \centering{\includegraphics[scale=0.6]{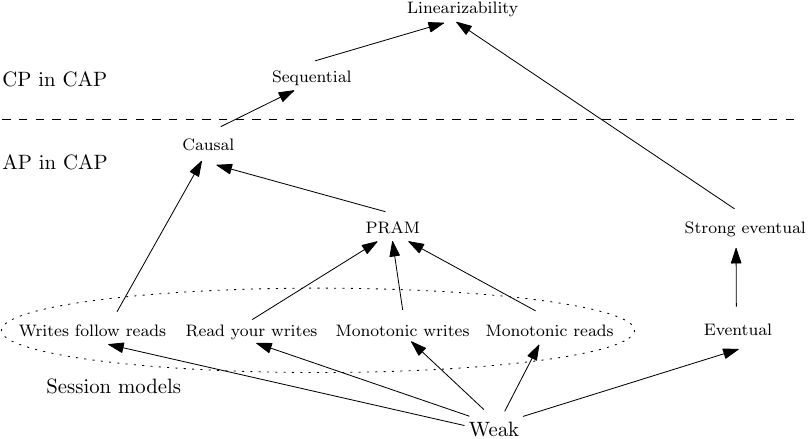}}
  \caption{Some consistency models for distributed systems.}
  \label{fig:consistency-models}
\end{figure}
\end{exclude}

\topic{Causal Consistency}

Causal Consistency (CC) allows processes to see different orders, as long as
they are consistent with a global \emph{happens-before} partial order.
Happens-before ($\rel{hb}$) contains the session order $\rel{so}$, which relates
operations from each process, and the \emph{visibility} order
$\rel{vis}$~\cite{DBLP:journals/ftpl/Burckhardt14}
(operation $a$ is visible to $b$, written $a \rel{vis} b$, if the effect of
$a$ is visible to the process performing $b$), being the transitive closure of
their union:
       \[ \rel{hb} \defeq (\rel{so} \union \rel{vis})^+ \]
and causal consistency essentially means that:
       \[ \rel{vis} = \rel{hb}, \]
i.e., all operations from the causal past are visible, with no missing
updates. Processes may possibly arbitrate operations in different total
orders, each compatible with the global visibility partial order, not
necessarily converging (causal consistency itself does not ensure
convergence).

\topic{Convergence}

The strongest guarantees that AP systems aim for are causal consistency
together with what has become known as \emph{strong eventual
consistency} (SEC)~\cite{DBLP:conf/sss/ShapiroPBZ11},
which guarantees that all updates will eventually become visible everywhere
(eventual visibility/delivery) and that processes that see the same set of
updates have equivalent state, regardless of the order in which they become
visible (\emph{strong convergence}).

We remark that the term SEC was an unfortunate choice of terminology, and the
source of some confusion, due to the use of the word \emph{strong}, usually
associated with strong consistency models, such as sequential consistency and
linearizability. Moreover, \emph{Eventual Consistency} (EC), as
originally introduced by~\textcite{DBLP:conf/pdis/TerryDPSTW94}
already includes the SEC guarantees. As described in that paper, EC systems
should include mechanisms to ensure two properties, on which EC relies:
\begin{itemize}
  \item \emph{total propagation}: each update is eventually propagated
    everywhere, by some anti-entropy mechanism (i.e., eventual delivery),
  \item \emph{consistent ordering}: non-commutative updates are applied in the
    same order everywhere, which implies the \emph{strong convergence}
    property of SEC.
\end{itemize}
These two properties mean, as described by~\textcite{DBLP:conf/sosp/TerryTPDSH95},
also addressing EC, that ``all servers eventually receive all Writes via
the pair-wise anti-entropy process and that two servers holding the same set
of Writes will have the same data contents''.
So, originally EC meant SEC, and a similar definition of EC was
also used by~\textcite{DBLP:journals/jpdc/RohJKL11}.
The informal meaning of EC ``eventual convergence when updates stop being
issued'', since it became popular due to~\textcite{DBLP:journals/cacm/Vogels09}, is
just a consequence of the above properties which EC systems ensure.
As originally introduced, what is ``eventual'' in EC is operation visibility
(delivery); replicas that have delivered the same set of updates have converged.

\begin{exclude}
Remark: It would be difficult to achieve convergence otherwise. If that was not the
case, \emph{when} would they converge? I.e., as result of processing
\emph{which} other events, namely if there were no more events to be acted
upon? So, SEC is not something that implies some extra cost over some
baseline, but the natural condition for EC systems, and the term SEC itself
should be deprecated in favor of using EC, as originally intended, together
with the terms eventual visibility and strong convergence for clarification.
\end{exclude}

\topic{Spatial scalability}

More than just being available (AP), EC systems aiming for CC and convergence,
can be performant, and usable at large spatial scales (very wide area), with
low operation response time. This is an essential property for interactive
systems, sometimes forgotten when thinking only of global throughput as a
measure of success, often done when evaluating CP systems.

Causal consistency can be achieved while being as fast as possible in terms of
physical limits (speed of light). We can say that these
systems exhibit ``mechanical sympathy'' regarding our universe, i.e., the
model suits the ``machine'', even at large spatial scales.  (The term
\emph{mechanical sympathy} was introduced in software by Martin Thomsom, to
refer to software being built with the understanding of how hardware works, so
that it can suit the hardware, and run efficiently in the underlying machine.)

On the contrary, linearizability can be said to be \emph{contra
naturam}. There is a wide mismatch between model (what it aims to guarantee) and
``machine'' (our universe). It would suit an Aristotelian universe, with
infinite light speed. In our universe, we must slow things down (delay responses)
to achieve it. The response time degrades with spatial span, becoming
impractical for very wide distances; e.g., linearizability would be completely unsuitable
for a system encompassing Earth and a future Mars colony.
Large spatial spans require specialized techniques or protocols, which
motivated research on DTNs (delay-tolerant networks) and OppNets (opportunistic
networks), for which CRDTs can be suitable.
\textcite{DBLP:journals/ppna/GuidecMN23} investigate the implementation of
CRDTs for OppNets.

\begin{exclude}
Similar observations about spatial scalability were made
by~\textcite{lipton1988pram}, when introducing a scalable memory (Pipelined
RAM), regarding \emph{coherency}, a weaker model than linearizability: ``Thus,
no matter how clever or complex a protocol is, if it implements a coherent
shared memory or CRAM, it must be `slow.' If a shared memory system must be
consistent, then it must take time proportional to $\tau_{global}$ for reading
and writing.'' (Where $\tau_{global}$ is the round-trip time across the
network.)
\end{exclude}

\subsection{Optimistic replication and CRDTs}

\topic{Optimistic replication}

Well before the CAP theorem was introduced, systems relaxing coordination and
providing availability were developed. This was the \emph{optimistic
replication}~\cite{DBLP:journals/csur/SaitoS05} approach. The main ingredients were:
  \begin{itemize}
    \item replicas are locally available for queries and updates;
    \item updates are propagated asynchronously, in the background,
      opportunistically.
  \end{itemize}

Thus, a total order of operations was not attempted: updates could become
known in different orders by different replicas.
This approach achieves both availability and low latency, but it poses a
problem: replicas may diverge, possibly forever. This problem was addressed
using two approaches, in two relevant early works:
  \begin{itemize}
    \item In Lazy replication~\cite{DBLP:journals/tocs/LadinLSG92}, relax
      ordering, but only for commutative operations.
    \item In Bayou~\cite{DBLP:conf/sosp/TerryTPDSH95}, have conflict
      detection, client-provided merge procedures, tentative writes for
      availability, but the same order at all replicas for committed writes
      (possibly undoing and reapplying if necessary) for eventual consistency.
  \end{itemize}

\topic{Conflict-free Replicated Data Types}

The introduction of Conflict-free Replicated Data Types
(CRDTs)~\cite{DBLP:conf/sss/ShapiroPBZ11} was an important step over previous
approaches to optimistic replication.
A CRDT is exposed as a standard data type, providing operations.
Each data type object is replicated and accessed locally by two kinds of
operations:
\begin{itemize}
\item mutator operations update state;
\item query operations look at the state and return a result.
\end{itemize}
Operations are always available, not depending on synchronization.
A CRDT object is highly available, even under partitions, with essentially
zero operation response time (only local computation).

As previous optimistic replication approaches, information is propagated
asynchronously. The main novelty is that conflicts are
dealt with semantically, making a replication-aware concurrent specification
part of the data type definition. This specification expresses how conflicts
are solved and, contrary to previous approaches, is not limited to commutative
operations. Conflict resolution is encapsulated by the data type, freeing
programmers from having to write application-specific ad hoc conflict
resolution code. (The effort becomes choosing the appropriate CRDTs.)

\topic{ADTs vs objects}

Evolution of object-oriented languages involved many
concepts, like abstract data type, parametric polymorphism, object, class,
inheritance and subtyping, as described in the classic
by~\textcite{DBLP:journals/csur/CardelliW85}, with tools such as existential
and bounded universal quantification. As well summarized
by~\textcite{DBLP:conf/oopsla/Cook09}, there are two main kinds of data
abstractions: ADTs and objects.

ADTs can be modeled as existential types, can access multiples instances, have
efficient binary operations, and different implementations cannot mix.
Objects can be modeled with recursive higher order functions, access only
the self state, binary operations are ``slow'' or even impossible, and multiple
implementations can be used together.

Most of the above is irrelevant to this paper, except for one aspect: ADT
implementations can have access to multiple instances (e.g., binary
operations, like multiply), while objects can only access the self state, with
other objects being only accessible through their interfaces.

For distributed systems, this means that only with full replication would
replicated ADTs be viable. In general, with only a subset of objects being
replicated in each node, we cannot rely on being able to access several
specific objects together. This is why CRDTs do not provide binary operations
involving two (or more) instances, but only operations on the ``self'' object.
Therefore, CRDTs normally are really ``Conflict-free Replicated Objects'',
which would have been a more suitable name.

\topic{Operation-based vs state-based approaches}
Concerning CRDT implementation (both the data type itself and the
propagation mechanism), there are two main approaches: operation-based and
state-based.

Operation-based approaches propagate information about operations to other
replicas, using a reliable messaging algorithm for propagation.
This normally needs some ordering guarantees, but weaker than a total order.
Normally causal delivery is chosen, to achieve the goal of having
causal consistency.
A special case is the \emph{pure} operation-based approach.

State-based approaches propagate replica states, as opposed to propagating
operations. A merge function is defined to be able to reconcile replica
states. State propagation is opportunistic, by ``background'' communication,
typically much less frequent than per-operation, to amortize the cost of
propagating full states.
An important variant, to make the propagation more incremental, is
the delta-state based approach, partially combining the advantages of both
approaches.

\section{Operation-based CRDTs}

The core concept of op-based (for short) CRDTs is to send operations,
not state, to other replicas, towards replica convergence.
So, when an update operation is invoked, in addition to being applied to the
replica where it was invoked, it is sent to all other replicas, asynchronously,
upon which they are applied at those replicas, when they arrive.
Query operations (that do not cause state changes) can make use of the local
state and be responded to immediately, causing no inter-replica messaging.

Because operations are not, in general, idempotent, it is essential that an
exactly-once messaging mechanism is used.
For some CRDTs no ordering guarantees at all would be needed for convergence.
But towards ensuring, in addition to convergence, causal consistency -- the
strongest possible consistency model while remaining available under
partitions -- a causal broadcast~\cite{DBLP:journals/tocs/BirmanJ87} mechanism
is normally adopted, making causally dependent operations become visible in
the correct order.

The essential improvement over previous attempts at optimistic
replication is the treatment of non-commutative operations.
Two concurrently invoked non-commutative operations
could arrive and be applied in different orders in different replicas, which
would lead to divergence. This is dealt with by sending more than just the
operation when ``broadcasting operations'' to other replicas.

\subsection{Execution model and concurrency semantics}

\topic{Standard execution model of operation-based CRDTs}

\begin{figure}
  \centering{\includegraphics[scale=0.9]{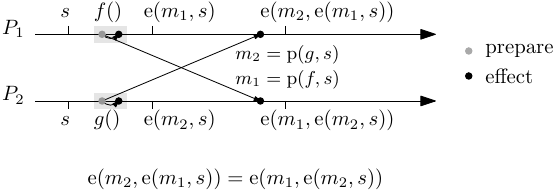}}
  \caption{Execution model for op-based CRDTs, stressing the commutativity of
  effect for concurrently invoked operations.}
  \label{fig:effect-commutativity}
\end{figure}

To achieve convergence even for data types containing non-commutative
operations, the execution model for op-based CRDTs, presented in
Figure~\ref{fig:effect-commutativity}, divides the execution of an update
operation in two phases: \emph{prepare} and \emph{effect}.

\begin{enumerate}
\item When an update operation is invoked, prepare is performed locally:
\begin{itemize}
  \item it looks at the state and the operation;
  \item it must have no side effects (on the abstract state);
  \item the result from prepare is disseminated with reliable causal broadcast.
\end{itemize}
\item Upon message delivery at each replica, effect is applied:
\begin{itemize}
  \item takes message (result from prepare) and state, and produces new state;
  \item it is designed to be commutative for concurrently invoked operations;
  \item it assumes immediate self-delivery on sender replica.
\end{itemize}
\end{enumerate}

Immediate self-delivery is important to ensure ``Read Your
Writes''~\cite{DBLP:conf/pdis/TerryDPSTW94}, making the state-change
immediately reflected locally and visible to subsequent query operations that
follow the update.

\topic{Reusing sequential data types with commutative operations}

Data types that only have commutative operations can be implemented as op-based
CRDTs trivially: the state is the same as for the sequential data type; prepare
returns the operation identifier and arguments; effect invokes the
corresponding sequential data type operation.
These data types respect the \emph{Principle of Permutation
Equivalence}~\cite{DBLP:conf/wdag/BieniusaZPSBBD12}: if all sequential
permutations of updates lead to the same state, then concurrent execution of
those operations should converge to that same state.
For such data types, not even FIFO order is needed for convergence, just
exactly-once delivery. Causal delivery is normally used to achieve causal
consistency.
Three examples are illustrated in Figure~\ref{fig:simple-CRDTs}: GCounter,
with only an increment update operation; a PNCounter, which may be negative,
having both increment and decrement; GSet, a grow-only set having just an add
update operation.

\begin{figure}
\begin{boxes}{3}
  \begin{tcolorbox}[title=GCounter]
    \begin{algo}{1}
  \cstate{
    $n : \nat = 0$
  }
  \query({$\af{value}() : \nat$}){
    \return $n$
  }
  \update({$\inc()$}){
    \prepare(){
      \return $\inc$
    }
    \effect({$\inc$}){
      $n \gets n + 1$
    }
  }
    \end{algo}
  \end{tcolorbox}
  \begin{tcolorbox}[title=PNCounter]
    \begin{algo}{1}
  \cstate{
    $v : \integer = 0$
  }
  \query({$\af{value}() : \integer$}){
    \return $v$
  }
  \update({$\inc()$}){
    \prepare(){
      \return $\inc$
    }
    \effect({$\inc$}){
      $v \gets v + 1$
    }
  }
  \update({$\dec()$}){
    \prepare(){
      \return $\dec$
    }
    \effect({$\dec$}){
      $v \gets v - 1$
    }
  }
    \end{algo}
  \end{tcolorbox}
  \begin{tcolorbox}[title=Grow-only set (GSet\typeparam{E})]
    \begin{algo}{1}
    \cstate{
      $s : \pow{E} = \emptyset$ \;
  }
  \query({$\af{elements}() : E$}){
    $\return\ s$ \;
  }
  \query({$\mathrlap{\af{contains}(e : E) : \bool}$}){
    $\return\ e \in s$ \;
  }
  \update({$\add(e : E)$}){
    \prepare(){
      $\return\ (\add, e)$
    }
    \effect({$(\add, e)$}){
      $s \gets s \union \{ e \}$ \;
    }
  }
  \end{algo}
  \end{tcolorbox}
  \end{boxes}
  \caption{Simple CRDTs with only commutative operations: GCounter, PNCounter
  and GSet. State and effect the same as for a sequential data type.}
  \label{fig:simple-CRDTs}
\end{figure}

\topic{CRDTs with non-commutative operations}

Where the CRDT approach becomes more interesting is for data types with
non-commutative operations.
An example is a set data type, having add and remove
operations. These are not commutative, as:
    \[ \add(v, \rmv(v, s)) \neq \rmv(v, \add(v, s)). \]

If the state were the same as for the sequential data type, and effect defined
as simply applying the corresponding operation, the possibility of different
delivery orders for concurrently invoked add and remove of the same element
would lead to divergence.

Therefore, the state must be more involved, and effect must be defined such
that it is commutative for concurrently invoked operations.
But not only convergence is relevant. How can we define what is supposed to
happen given two concurrently invoked add and remove? I.e., how can we define
the data type semantics for such CRDTs?

\topic{Defining concurrent semantics}

A first design criteria for CRDT semantics is preserving the sequential
semantics of the original data type. I.e., under a sequential execution the
CRDT should produce the same outcome as the corresponding sequential data type.
Then, we must define how to handle conflicts for concurrently invoked
operations. In the set example, given concurrent add and remove of the same
element, we want to define which will ``win''. We have several possibilities:
  \begin{itemize}
    \item add wins;
    \item remove wins;
    \item last-writer-wins (LWW), using a totally ordered arbitration.
  \end{itemize}

For CRDTs, in general, the outcome may not be equivalent to some sequential
execution. Data type semantics are usually defined resorting to the causal
past, i.e., the result from a query depends on the set of update operations
that are visible to it, and their partial order under \emph{happens-before}.

\subsection{Observed-cancel CRDTs}

\topic{Observed-cancel semantics}

We may define different CRDTs, with different semantics, for each sequential
data type (e.g., for a set). A particularly interesting concept, for choosing
in which way a conflict between two operations is handled, is what can be
called \emph{observed-cancel} semantics. Essentially, ``cancel
observed (visible) operations, as if they were never issued''.

Using causal delivery, as usual for op-based CRDTs, this means that an
operation which cancels another will cancel the operations in its causal
past, but not the ones concurrently issued. This is the most appealing choice
because it makes an operation act upon a closed, well defined set of other
operations which have already taken effect on the state. It will not
``blindly discard'' updates not yet seen, and that could not have been
accounted for yet. (Such updates will eventually become visible and can always
be canceled subsequently.) This prevents undesired ``lost
updates'' which CRDTs aim to avoid.
Two examples are the observed-remove set~\cite{DBLP:journals/corr/abs-1210-3368}:
  \begin{itemize}
    \item has add and remove operations;
    \item remove only cancels the adds visible to it;
    \item concurrent adds will not be canceled and will ``win'';
  \end{itemize}
and the observed-reset counter~\cite{DBLP:conf/eurosys/WeidnerA22}:
  \begin{itemize}
    \item has increment and reset;
    \item a reset cancels the observed increments.
  \end{itemize}

The appeal of observed-cancel semantics can be seen in the observed-reset
counter: it allows a sample-and-reset pattern, where a process
periodically samples the counter value and resets it. This allows grouping
increments in a sequence, without losing any increment, even
the ones concurrently issued, which will be accounted for in the next
sample-and-reset. An alternative semantics where reset cancels
concurrent increments would not allow accurate accounting to be achieved.

\topic{Observed-remove set}

A well known example is precisely the observed-remove
set, also called add-wins set, because adds win over concurrent removes.
Several different implementations were presented in the literature, from very
naive, to ``optimized''~\cite{DBLP:journals/corr/abs-1210-3368}.
An even more optimized version, not previously published, is presented in
Figure~\ref{fig:observed-remove-set}.

Many CRDTs start indeed from a naive version, being further optimized along
time. The observed-remove set is a good example of possible improvements. A
vanilla, naive, version is shown in Figure~\ref{fig:naive-observed-remove-set}.
In this CRDT, the state is a set of pairs and an element $e$ is considered to
belong to the set if there is a pair $(e, u)$, for some unique identifier $u$,
in the set of pairs.
This version has some open issues and several possible improvements.

The first issue is that it assumes the generation of unique ids,
but does not specify how to do so. This is a frequent need, and can be
achieved by using unique replica ids and a counter per replica, incremented at
each operation. Unique ids are obtained as pairs (replica id, counter);
this is what we call a ``dot'' (from Dotted Version
Vectors~\cite{DBLP:journals/corr/abs-1011-5808}).
The second issue is that, being the state a set of pairs, most operations need
set traversal, which is inefficient. The solution is to use a map from
elements to sets of ids.
A third issue is that adds keep accumulating state, adding a new pair to
the ones from previous adds of the same element. An improvement is to replace
the current pairs, for the given element, with a single pair for the new unique id.
A fourth issue is that prepare for remove sends the element being removed
repeated in each pair. The improvement is to collect the set of ids
separately.

\begin{figure}
  \begin{algo}{2}
  \cstate{
    $s : \pow{E \times \text{\ldots}} = \emptyset$ \;
  }
  \query({$\af{elements}() : E$}){
    \return $\{ e | (e, \_) \in s \}$
  }
  \query({$\af{contains}(e : E) : \bool$}){
    \return $\exists x \cdot (e, x) \in s$
  }
  \BlankLine
  \update({$\add(e : E)$}){
    \prepare(){
      \kwlet $u = [\text{some unique id}]$ \;
      \return $(\add, e, u)$
    }
    \effect({$(\add, e, u)$}){
      $s \gets s \union \{(e, u)\}$
    }
  }
  \update({$\af{remove}(e : E)$}){
    \prepare(){
      \kwlet $r = \{ (x, u) \in s | x = e\}$ \;
      \return $(\af{remove}, r)$
    }
    \effect({$(\af{remove}, r)$}){
      $s \gets s \setminus r$
    }
  }
\end{algo}
  \caption{Op-based observed-remove set, ORSet\typeparam{E}, naive
  implementation.}
  \label{fig:naive-observed-remove-set}
\end{figure}

\begin{figure}
  \begin{algo}{2}
  \types{
    $\ids$, set of replica identifiers
  }
  \parameters{
    $i \in \ids$, replica identifier
  }
  \cstate{
    $m : E \pfunc \pow{\ids \times \nat} = \emptyset$ \;
    $c : \nat = 0$, auxiliary state
  }
  \query({$\af{elements}() : E$}){
    \return $\dom m$
  }
  \query({$\af{contains}(e : E) : \bool$}){
    \return $m[e] \neq \emptyset$
  }
  \update({$\add(e : E)$}){
    \prepare(){
      $c \gets c + 1$ \;
      \return $(\add, e, (i, c), m[e])$
    }
    \effect({$(\add, e, d, r)$}){
      $m[e] \gets m[e] \setminus r \union \{d\}$
    }
  }
  \update({$\af{remove}(e : E)$}){
    \prepare(){
      \return $(\af{remove}, e, m[e])$
    }
    \effect({$(\af{remove}, e, r)$}){
      $m[e] \gets m[e] \setminus r$
    }
  }
  \end{algo}
  \caption{Op-based observed-remove set ORSet\typeparam{E}, optimized
  implementation. Algorithm for replica $i$.}
  \label{fig:observed-remove-set}
\end{figure}

The optimized implementation, in Figure~\ref{fig:observed-remove-set},
contains all the above improvements. It assumes a unique replica identifier $i$ in the
set $\ids$ of possible replica identifiers, and assumes the standard op-based
execution model, using causal delivery.
Each replica has a counter, incremented per add, which allows, together with
the replica id, to generate unique ids.
It must be noticed that this counter is auxiliary state, not part of the CRDT state
used in queries or effect. This auxiliary state does not converge and it can be updated
in prepare. This allows not respecting the classic rule that prepare must be free
from side-effects, and obtaining a better CRDT implementation. Currently
published versions which do not make this distinction are less elegant.

The state is a map from elements in the set to sets of operation identifiers.
Here we assume that the map stores only non-empty sets, implicitly returning
$\emptyset$ for unmapped keys.
Prepare returns a tuple with operation, argument, unique id in the
case of an add, and the set of ids which the map holds, for the
given element. When effect is applied for remove, it subtracts the set of
ids sent from the ones in the map, for the corresponding element. This
has the desired outcome of removing the adds that have been observed at the
replica where the remove was invoked, at the time it was invoked. Concurrently
issued adds will ``survive''. For add, effect adds the newly generated id and
subtracts the set of ids present when the add was issued (similar to remove),
as the new id makes the others redundant.

This CRDT is more optimized than the one previously
published~\cite{DBLP:journals/corr/abs-1210-3368}: it avoids computation cost
by organizing entries in a map; avoids sending the element repeatedly in a
remove; and removes all observed identifiers for the element in an add, while
the previously published only discards entries of previous adds from the same
source. This exemplifies how, even for a relatively simple CRDT, many
different versions with subtle variations may exist. This example also
shows the role of commutativity: not only the operations themselves
(add/remove) are not commutative, but the effect of add/remove is also not
commutative: only the effect of concurrently issued add/remove is so.

\subsection{Beyond sequential semantics and API}

\topic{Lack of equivalence to sequential executions}

CRDTs aim to preserve the sequential semantics for sequential executions, but
what concerns concurrent invocations not always is it possible to
achieve such equivalence. In some cases the CRDT:
  \begin{itemize}
    \item has behavior not possible by any sequential execution;
    \item the interface itself is different from the sequential data type.
  \end{itemize}

The observed-remove set allows executions which are not equivalent to any
sequential execution, considering the corresponding sequential data type
semantics (of a set). This is easily seen by a run with two replicas,
involving two elements $a$ and $b$, where concurrently each replica adds an
element and removes the other, i.e.,
$\add(a);\af{remove}(b) \parallel \add(b);\af{remove}(a)$.
By the observed-remove set semantics, upon convergence both elements will be
in the set, which is impossible in a sequential execution, as some remove
will be the last operation.

Moreover, there are CRDTs for which even the interface itself was changed from
the corresponding sequential data type. The most well known example with a modified
interface is the multi-value register, made popular by the
Dynamo~\cite{DBLP:conf/sosp/DeCandiaHJKLPSVV07} key-value store from Amazon.

\topic{Multi-value register}

The multi-value register keeps the set of the most recent concurrent writes.
A read returns that set of values, and a write overwrites that set, in the
current replica into a singleton. A multi-value register implementation, using
the same technique of generating unique ids as for the
observed-remove set is presented in Figure~\ref{fig:multi-value-register}.

\begin{figure}
  \begin{algo}{2}
  \types{
    $\ids$, set of replica identifiers
  }
  \parameters{
    $i \in \ids$, replica identifier
  }
  \cstate{
    $s : \pow{E \times (\ids \times \nat)} = \emptyset$ \;
    $c : \nat = 0$, auxiliary state
  }
  \query({$\af{read}() : \pow{E}$}){
    $\return\ \{ e | (e, \_) \in s \}$ \;
  }
  \update({$\af{write}(e : E)$}){
    \prepare(){
      $c \gets c + 1$ \;
      \kwlet $r = \{d | (\_, d) \in s \}$ \;
      $\return\ (\af{write}, (e, (i, c)), r)$
    }
    \effect({$(\af{write}, v, r)$}){
      $s \gets \{(e, d) \in s | d \not\in r \} \union \{v\}$
    }
  }
\end{algo}
  \caption{Op-based multi-value register, MVReg\typeparam{E}. Algorithm for replica $i$.}
  \label{fig:multi-value-register}
\end{figure}

The state is a set of pairs, each with the written value and corresponding
unique id. Prepare returns a tuple with: the operation, a pair with
value and unique id, and the set of ids in the state. When effect is
applied, it keeps the pairs in the state with id not present in the set of ids in the
message, and adds the new pair. This has the desired outcome of removing
writes made obsolete, only preserving the most recent concurrent writes. At
the replica where the write is invoked only a singleton will remain.

\section{Pure operation-based CRDTs}

\topic{Pure op-based CRDTs}

Pure op-based CRDTs~\cite{DBLP:conf/dais/BaqueroAS14} are a subset of general
op-based CRDTs, restricted to the essence of ``send only operations''.
Pure op-based CRDTs use the same prepare-effect execution model. What defines
them is the restriction that:
\begin{itemize}
\item Prepare simply returns the operation (including arguments), ignoring
  current state.
\end{itemize}
Given operation $o$ and state $s$:
\[
\prepare(o, s) = o.
\]

This is unlike the general op-based approach, which can depart too much from
the op-based spirit, allowing implementations that are, for all practical
purposes, state-based: We can pick any state-based CRDT, define prepare as
applying the operation and returning the resulting state, and define effect as
merging states.
However, pure op-based CRDTs may be less efficient
than the general op-based approach. The pure-op based
approach is relevant as an important point in the design space.

\topic{Pure implementations of commutative data types}

For data types with only commutative operations, where for any operations
$f$ and $g$, and state $s$:
\[
f(g(s)) = g(f(s)),
\]
operations can be applied in any order, producing the same result.
As discussed in the general op-based model, the CRDT specification can be
based on the sequential specification, with a trivial implementation, defining
effect as simply applying the operation:
\[
\effect(o, s) = o(s).
\]
This applies to CRDTs such as GCounter, PNCounter, and GSet, described before,
which fit the pure-op model.
Again, the challenge lies on non-commutative operations: how can we obtain
pure op-based implementations for non-commutative data types?  The solution is
to use an augmented form of causal broadcast, called \emph{tagged causal
broadcast}, which provides knowledge about happens-before and about
\emph{causal stability}, which we describe below.


\subsection{Tagged Causal Broadcast and Causal Stability}

\topic{Tagged Causal Broadcast (TCB)}

Causal broadcast middleware already manages causality information, but
normal APIs do not expose it to clients.
\emph{Tagged Causal Broadcast}~\cite{DBLP:conf/dais/BaqueroAS14} provides:
\begin{itemize}
\item a partial order on messages, in terms of an end-to-end happens-before;
\item information about \emph{causal stability} of messages, as we define
  below.
\end{itemize}

The TCB API defines delivery as providing, together with the message
itself, a timestamp corresponding to a partially ordered logical clock value
which reflects \emph{happens-before}.
This timestamp can be used in the implementation of the CRDT to distinguish
between causally related and concurrently invoked operations, and implement
the desired semantics.

\topic{Causal Stability}

We define causal stability as: \emph{message with timestamp $t$ is
\emph{causally stable} at node $i$ when all messages subsequently delivered at
$i$ will have timestamp $t' \ge t$}.
So, a message is causally stable when no more concurrent messages will be
delivered. This is different from classic multicast/message
stability~\cite{DBLP:journals/tocs/BirmanSS91}. 

A multicast is stable when it has been received by all nodes. Multicast stability
is used internally by the messaging middleware, being useful for garbage
collection when ensuring fault tolerance: once a message has been
delivered at some node and becomes stable, that node can discard it.

While classic multicast stability regards messages being received, being an
implementation aspect, causal stability is a property involving delivery,
visible to TCB clients. Also, while multicast stability is a global property,
causal stability is a per-node property: some message may be causally stable
in some node but not in others.  Causal stability is stronger: a message only
becomes causally stable in some node when it has become stable.

\topic{The many faces of stability}

``Stability'' has been a much overloaded expression in distributed systems.
Table~\ref{tab:stability} shows some terms where the word stability is
used, and their rough meaning. Causal stability has not been properly
recognized, being sometimes confused with message stability. Although
the concept itself was not new when pure op-based CRDTs were introduced,
coining the term ``causal stability'' will help avoid some confusion,
specially in contexts where both may coexist, such as when describing a
system involving pure op-based CRDTs, which rely on causal stability, making use
of a TCB middleware, whose implementation may resort to multicast/message stability.

\begin{table}
  \caption{Some usages of ``stability'' and their meaning.}
  \small
  \begin{center}
    \begin{tabular}{@{}lll@{}}
  \toprule
    Term & Provenance & Meaning \\
  \midrule
    Self stabilization & \textcite{DBLP:journals/cacm/Dijkstra74} & returns to
    valid state \\ 
    Stable storage & \textcite{lampson1979crash} & durable storage, survives crashes \\
    Multicast stability & \textcite{DBLP:journals/tocs/BirmanSS91} & message was
    received by all nodes \\
    Write stability & \textcite{DBLP:conf/sosp/TerryTPDSH95} & a tentative write commits \\
    Causal stability & \textcite{DBLP:conf/dais/BaqueroAS14} & concurrent messages
    were delivered \\
  \bottomrule
  \end{tabular}
  \end{center}
  \label{tab:stability}
\end{table}

\topic{Distributed algorithm for pure op-based CRDTs}

The distributed algorithm for pure op-based CRDTs becomes simply reacting to
the TCB middleware callbacks and invoking either prepare, effect, or a
$\af{stable}$ function to make use of causal stability information, as shown
in Figure~\ref{fig:op-based-algo}.

\begin{figure}
\begin{algorithm}[H]
\DontPrintSemicolon
\SetKwBlock{state}{state:}{}
\SetKwBlock{on}{on}{}
\state{
  $s \in S$
}
\on({$\operation(o)$:}){
  $\tcbcast(\prepare(o, s))$
}
\on({$\tcdeliver(m, t)$:}){
  $s \gets \effect(m, t, s)$
}
\on({$\tcstable(t)$:}){
  $s \gets \stable(t, s)$
}
\end{algorithm}
  \caption{Distributed algorithm for pure op-based CRDTs, making use of a TCB
  middleware.}
  \label{fig:op-based-algo}
\end{figure}

\subsection{Resorting to a partially ordered Log}

\topic{Naive PO-Log based implementations}

A starting point for implementing pure op-based CRDTs is
making the state a \emph{\polog}: partially ordered log of operations.
The \polog can be implemented as a map from timestamps to operations.
Effect simply adds an entry $(t,o)$ delivered by TCB to the \polog.
This makes both prepare and effect to have a universal definition.
Only queries are data type dependent, being defined over the \polog.
This approach is shown in Figure~\ref{fig:polog-impl}.
It is very naive, but a starting point for subsequent optimization.

\begin{figure}
\begin{eqnarray*}
S = T \map O &&
s^0 = \{\}  \\
\prepare(o, s) & = & o \\
\effect(o, t, s) & = &  s \cup \{(t,o)\} \\
\eval(q, s ) & = & \textnormal{[data type specific query function over \polog]}
\end{eqnarray*}
  \caption{\polog based implementation for pure op-based CRDTs.}
  \label{fig:polog-impl}
\end{figure}

\topic{\polog based observed-remove (add-wins) set}

An example of a \polog based CRDT, an observed-remove set is shown
in Figure~\ref{fig:polog-add-wins}. Prepare and effect have the universal
definition shown before. Only query is data type specific.
The query mimics the specification over the partial order of operations: an
element is considered to be in the set if there exists an add for that element
not canceled by a remove in its causal future.
Essentially, this implementation is a naive runnable specification. But it shows
the role of partial-order based concurrent specifications in the design of
CRDTs~\cite{burckhardt2013understanding,DBLP:conf/popl/BurckhardtGYZ14}, and
how it fits directly the pure op-based model, as opposed to the totally
ordered history of operations used for sequential specifications.

\begin{figure}
\begin{eqnarray*}
S = T \map O &&
s^0 = \{\}  \\
\prepare(o, s) & = & o
\qquad\qquad\qquad (o \textnormal{ either } [\add, v] \textnormal{ or
} [\rmv, v])\\
\effect(o, t, s) & = &  s \cup \{(t,o)\} \\
  \eval(\af{elements}, s ) & = & \{ v | (t, [\add,v])
\in s \ \land \not\exists (t',[\rmv,v]) \in s \cdot t < t'\}
\end{eqnarray*}
  \caption{\polog based observed-remove (add-wins) set CRDT.}
  \label{fig:polog-add-wins}
\end{figure}

\topic{Semantically based \polog compaction}

\polog based CRDTs as described would be very inefficient, and need
optimizations to be actually usable. The idea is to avoid \polog growth,
making it compact, by keeping the smallest number of items that
produce equivalent results when queries are performed.
The compaction is data type specific, according to the specification, and
makes use of the causality related data provided by the TCB middleware:
\begin{itemize}
  \item Causality: to prune the \polog after effect is performed, i.e., after
    operation delivery;
  \item Causal stability: to discard timestamps for operations that become
    causally stable and in some cases to remove operations that become
    redundant after causal stability.
\end{itemize}

Causality information can be exploited to achieve \polog compaction by
redefining $\effect$ to use a data type specific $\af{obsolete}$ relation
between timestamps:
\[
\effect(o, t, s) \defeq
\{ x \in s | \lnot \obsolete(x, (t,o)) \} \union
\{(t,o) | x \in s \implies \lnot \obsolete((t,o), x) \}.
\]
When a new pair $(t, o)$ is delivered, $\effect$ discards from
the \polog all elements $x$ such that $\obsolete(x, (t,o))$ holds. Moreover, the
delivered pair $(t,o)$ is only inserted into the \polog if it is not itself
redundant, i.e., if $\obsolete((t,o), x)$ is false for any $x$ in the \polog.
The $\obsolete$ relation is not restricted to being a partial-order to allow,
e.g., a newly arrived operation to discard others in the \polog without
necessarily being itself added.

Regarding exploiting causal stability, the basic improvement is to replace
timestamps by $\bot$, when they becomes causally stable, as they will be
compared as in the past of new operations that will arrive, saving space.
For some data types we can also remove some causally stable operations
themselves, if they become redundant considering the data type semantics.
We thus define $\stable$ as:
\[
  \stable(t, s) \defeq \stabilize(t, s)[ (\bot,o) / (t,o)], \\
\]
where $\stabilize$ is a data type specific function which discards operations
from the \polog that become redundant upon causal stability.

Commonly, $\stabilize$ is the identity function, but it can be quite involved. A
detailed description of how causal stability can be exploited is beyond the scope of
this paper. Details can be found
in~\textcite{DBLP:conf/dais/BaqueroAS14,DBLP:journals/corr/abs-1710-04469}.
Just to give an example, in a remove-wins set, a remove
wins over a concurrent add of the same element and must be kept in the \polog
for some time, in case a concurrent add shows up. But once the remove becomes
causally stable it can be discarded in some cases.

\topic{\polog based observed-remove set with \polog compaction}

To exemplify how causality information can be used to perform \polog
compaction, for an ORSet it is true that:
\begin{itemize}
\item a subsequent $\add$ obsoletes a previous $\add$ of the same value;
\item a subsequent $\af{remove}$ obsoletes a previous $\add$ of the same value;
\item a $\af{remove}$ is made obsolete by any other timestamped operation:
  after having made other operations obsolete, the $\af{remove}$ itself
    becomes redundant.
\end{itemize}

So, the $\af{obsolete}$ relation for the ORSet can be defined as:
\begin{eqnarray*}
\obsolete((t, [\add, v]), (t', [\add, v'])) &=&  t < t' \land v = v' \\
  \obsolete((t, [\add, v]), (t', [\af{remove}, v'])) &=&  t < t' \land v = v' \\
  \obsolete((t, [\af{remove}, v]), x) &=& \true
\end{eqnarray*}
The compacted \polog not also saves space but also allows a substantially
simpler eval:
\begin{eqnarray*}
  \eval(\af{elements}, s ) & = & \{ v | (t, [\add,v]) \in s \}
\end{eqnarray*}

\topic{On partition tolerance}

A possible criticism that can be made regarding the use of causal stability is
that operations stop becoming stable under partitions. We observe that, while
for strongly consistent (CP) systems a partition causes unavailability, in
pure op-based CRDTs a partition only impacts state size, which may grow while
partitioned, but does not impact availability itself. If the partition does
not take long, it may not cause much harm and the system will recover. We also
note that long partitions will be a problem in general for op-based (not only
pure op-based) CRDTs, due to the need for buffering messages by the reliable
messaging middleware. If long partitions or disconnected operation is the
norm, state-based CRDTs are preferable.

\section{State-based CRDTs}

\topic{State-based propagation and conflict resolution}

The state-based approach propagates replica states to other replicas.
If the self state and a received state conflict, they must be merged
(reconciled) into a new state reflecting the conflict resolution.
In classic optimistic replication this was performed in some ad hoc way,
through a user-defined merge procedure.

Contrary to the operation-based approach, where each operation is immediately
propagated, propagating states, which can be large, is performed much less
frequently.  Also, as opposed to the usual causal broadcast to a well known
group of replicas in the operation-based approach, state-based propagation can
have many forms, e.g., through an epidemic
propagation~\cite{DBLP:conf/podc/DemersGHIL87} to an unknown set of
participants.
Given that states can be propagated in many different ways and arrive at a
replica in different orders, conflict resolution (merge) should be:
\begin{itemize}
  \item deterministic: result as a function of inputs (current and received
    state);
  \item obviously a commutative function;
  \item associative: to give the same result when merging in different orders;
  \item idempotent: merging equal states produces that state;
  \item merging with an older state produces the current state;
  \item monotonic: merging with a ``newer'' state produces a ``newer'' state.
\end{itemize}

The solution is to adopt for the replica state the mathematical
concept of \emph{join-semilattices}~\cite{birkhoff1940}. We now present a
very brief introduction to the concepts of order and lattices relevant for
CRDTs.

\subsection{Lattices and Order}

\topic{Partially ordered sets (posets)}

A partially ordered set (poset), has a binary relation $\pleq$ which is:
  \begin{itemize}
    \item (reflexive) $p \pleq p$;
    \item (transitive) $o \pleq p \land p \pleq q \implies o \pleq q$;
    \item (anti-symmetric) $p \pleq q \land q \pleq p \implies p = q$.
  \end{itemize}
Two unordered elements are called concurrent:
  \begin{itemize}
    \item (concurrent) $p \parallel q \iff \neg (p \pleq q \lor q \pleq p).$
  \end{itemize}
Some posets have a \emph{bottom} ($\bot$), an element smaller than any other:
  \[ \forall p \in P \cdot \bot \pleq p. \]

\topic{Join-semilattices}

An upper bound of some subset $S$ of some poset $P$ is some $u$ in $P$ greater
or equal than any element in $S$:
  \[\forall s \in S \cdot s \pleq u. \]
If the \emph{least upper bound} of $S$ exists (an upper bound smaller than any
other), it is called the \emph{join} of $S$, denoted $\bigjoin S$.
For two elements, $\bigjoin \{p, q \}$ is denoted by $p \join q$.
A poset $P$ is a join-semilattice if $p \join q$ is defined for any two
elements $p$ and $q$ in $P$.
The join operator has the following properties:
  \begin{itemize}
    \item (idempotent) $p \join p = p$;
    \item (commutative) $p \join q = q \join p$;
    \item (associative) $o \join (p \join q) = (o \join p) \join q$.
  \end{itemize}

\topic{Examples and non-examples of join-semilattices}

Some examples of join-semilattices are:
  \begin{itemize}
    \item natural numbers $\nat$: ${\pleq} = {\leq}$; $\join = \max$; $\bot = 0$;
    \item booleans $\bool$: $\false \ple \true$; $\join = {\lor}$; $\bot = \false$;
    \item any totally ordered set (called a \emph{chain}).
  \end{itemize}
Some posets which are not join-semilattices are:
  \begin{itemize}
    \item unordered posets (called \emph{antichains}), where all elements are
      incomparable;
    \item strings under prefix ordering (e.g., $small \pleq smallest \parallel
      smaller$), where all concurrent elements are not joinable.
  \end{itemize}

\topic{Lattice compositions}

An advantage of adopting (join-semi)lattices for the replica state is that we
can use standard lattice composition constructs for obtaining complex states
from simpler states. Some examples are given in
Figure~\ref{fig:lattice-compositions}:
the (Cartesian) product of two lattices, where join is performed
component-wise; the lexicographic product of $A$ and
$B$, when elements are lexicographic pairs, compared first by the left-side
component and only by the right-side one to break ties, being defined when $B$
has a bottom or $A$ is a chain; the powerset of any set, being join the
set union; and the function space from any set $A$ to lattice $B$, where both
comparison and join are performed pointwise. Maps, where missing keys
implicitly yield bottom can be considered a special case of functions, being
specially useful. These and other examples are presented
by~\textcite{DBLP:journals/eatcs/BaqueroACF17}.

\begin{figure}
  \newcommand\fsize{\small}
  \begin{boxes}{2}
    \begin{tcolorbox}[title=Cartesian product $A \times B$]
      \fsize
     \begin{center}$\begin{aligned}
      (a_1, b_1) \pleq (a_2, b_2) &= a_1 \pleq a_2 \land b_1 \pleq b_2 \\
      (a_1, b_1) \join (a_2, b_2) &= (a_1 \join a_2, b_1 \join b_2)
     \end{aligned}$\end{center}
      \sep
      (join-semilattices $A$ and $B$)
    \end{tcolorbox}
    \begin{tcolorbox}[title=Lexicographic product $A \boxtimes B$]
      \fsize
     \begin{center}$\begin{aligned}
    (a_1,b_1) \pleq (a_2,b_2) &= a_1 \ple a_2 \lor
    (a_1 = a_2 \land b_1 \pleq b_2) \\
      (a_1,b_1) \join (a_2,b_2) &= 
  \begin{cases}
    (a_1, b_1) & \text{if}\ a_2 \ple a_1\\      
    (a_2, b_2) & \text{if}\ a_1 \ple a_2\\
    (a_1, b_1 \join b_2) & \text{if}\ a_1 = a_2\\      
    (a_1 \join a_2, \bot) & \text{if}\ a_1 \parallel a_2  
  \end{cases}
     \end{aligned}$\end{center}
      \sep
      (when $B$ has a bottom or $A$ is a chain)
    \end{tcolorbox}
    \begin{tcolorbox}[title=Powerset $\pow{S}$]
      \fsize
     \begin{center}$\begin{aligned}
    {\pleq} &= {\subseteq} \\
      \join &= \union 
     \end{aligned}$\end{center}
      \sep
      (set $S$)
    \end{tcolorbox}
    \begin{tcolorbox}[title=Function space $A \to B$]
      \fsize
     \begin{center}$\begin{aligned}
    f \pleq g &= \forall x \in A \cdot f(x) \pleq g(x) \\
      (f \join g)(x) &= f(x) \join g(x) 
     \end{aligned}$\end{center}
      \sep
      (set $A$ to join-semilattice $B$)\\
      \sep
      (map $K \pfunc V$, $V$ with bottom, where missing keys yield
      bottom is specially useful)
    \end{tcolorbox}
  \end{boxes}
  \caption{Some classic lattice compositions.}
  \label{fig:lattice-compositions}
\end{figure}

\subsection{Basics of State-based CRDTs}

\topic{State-based CRDTs}

Like for op-based CRDTs, in state-based CRDTs update operations are applied
locally, and queries can be answered from the local state.
The essential difference is how knowledge about operations is propagated.
The propagation is indirect, through states: an operation
updates the local state; from time to time replicas send their full state to other
replicas.

For this strategy to work, the essential aspect of state-based CRDTs is making
the state to be a join-semilattice. Replicas merge the received state using
the join operator $\join$. The properties of join give state-based CRDTs very
good fault tolerance in terms of messaging faults: they work in unreliable
networks subject to message loss, duplication, and reordering.
Also, sending the full state ensures causal consistency, as it ensures a
transitive propagation of the causal past.

\topic{Update operations and state mutators}

In the state-based model, update operations invoke a \emph{state mutator},
which returns the new state. A state mutator must be an \emph{inflation}:
  \begin{itemize}
    \item (inflation) $x \pleq f(x)$;
    \item (strict inflation) $x \ple f(x)$.
  \end{itemize}
This, together with join being used for merging states, gives the essential property that replica
states are always ``going up'' in the partial order, i.e., state evolves
monotonically:
\begin{itemize}
  \item when updates are locally invoked;
  \item when a remote state is received and merged through join.
\end{itemize}

Monotonic evolution of state is essential, and what gives nice fault tolerance
properties: a newer state always subsumes an older state; if a message is
lost, a newer one will subsume it; an old message can arrive as a duplicate
causing no harm.

\topic{Inflations vs monotonic functions}

\begin{figure}
\begin{center}
  \includegraphics[scale=0.7]{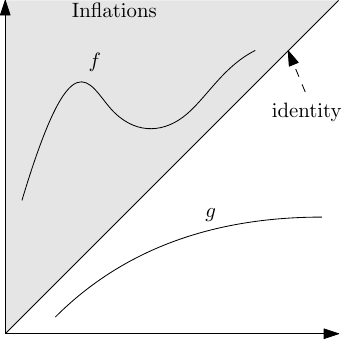}
\end{center}
  \caption{Inflations vs monotonic functions.}
  \label{fig:inflations}
\end{figure}

Sometimes there is some confusion about monotonicity.
Erroneously, many places say mutators need to be monotonic, when in fact 
mutators need to be inflations.
Being monotonic is not necessary nor sufficient, as illustrated in
Figure~\ref{fig:inflations}: $f$ is an inflation but not monotonic and
$g$ is monotonic but not an inflation.
Decrement ($f(x) = x - 1$) is monotonic but not an inflation.
What is monotonic is state evolution over time, as result of applying mutators
or join.

\topic{Reusing sequential data types}

As for op-based CRDTs, for some data types we can have trivial state-based
CRDTs, with state and operations being the same as the corresponding
sequential data type. But for state-based CRDTs this is considerably more difficult.
It is necessary that:
\begin{itemize}
  \item the state is already a join-semilattice (can be ordered, if not
    already, to be one)
  \item update operations are inflations.
  \item update operations are idempotent;
\end{itemize}

One may think that another condition for reuse would be that that the
sequential data type should only have commutative operations, as for op-based CRDTs.
This is not true. A counter-example is a sequential data type, presented in
Figure~\ref{fig:seq-advancer}, whose state is a
map from some unspecified set of keys to integer values; with a single
operation $\af{advance}(k)$, which makes the corresponding key have a value at
least one more than any other key, by increasing the value by the minimum
needed (possibly 0); and with a single query $\af{ahead}(k)$, which returns
the set of keys with the maximum value (an empty set for the initial state or
a singleton afterwards, for sequential executions).

\begin{figure}
\begin{eqnarray*}
  \af{Advancer}\typeparam{E} &=& E \pfunc \nat \\
  s^0 &=& \{\} \\
  \af{advance}(e,s) & = & s\{e \mapsto
  \max\{s[e], 1 + \max\{v | (k,v) \in s \land k \not= e\}\}\} \\
  \af{ahead}(s) & = & \{k | (k,v) \in s \land
  v = \max\{x | (\_,x) \in s\}\}
\end{eqnarray*}
  \caption{Sequential ``advancer'' data type, reusable as state-based CRDT.}
\label{fig:seq-advancer}
\end{figure}

The update operation is not commutative as, starting from the initial state
(empty map):
\[
  \af{advance}(a); \af{advance}(b) \quad \textnormal{leads to} \quad
  \{ a \mapsto 1, b \mapsto 2 \},
\]
and
\[
  \af{advance}(b); \af{advance}(a) \quad \textnormal{leads to} \quad
  \{ a \mapsto 2, b \mapsto 1 \}.
\]

But the original state (a map from some set to integers) is already a lattice,
with join as the pointwise maximum, and the update operation is an inflation
and idempotent.
We can simply reuse the sequential data type state and operations and
the original lattice join for merge. Concurrent execution of the operations
above would be:
\[
  \af{advance}(a) \concurrent \af{advance}(b) \quad \textnormal{leads to} \quad
  \{ a \mapsto 1, b \mapsto 1 \}.
\]

This example is interesting as it shows that for this data type concurrent
executions can lead to states unreachable by any sequential execution. But
this CRDT reuses the sequential data type and preserves its sequential
semantics, while converging and leading to reasonable outcomes for
concurrent executions: concurrent invocations of advance may lead to ties,
unlike sequential executions, which is a reasonable generalization of the
sequential behavior.

In this example reuse was possible because the update operation, even though
not commutative, was an inflation and idempotent. All updates being inflations
is obviously necessary for reuse. Having to be idempotent is also easy to
show, given the desire for permutation equivalence. Given a non-idempotent
inflation $o$, the desired outcome for $o(s) \concurrent o(s)$ is $o(o(s))$, but 
\[ o(s) \concurrent o(s) \quad \textnormal{converges to} \quad  o(s) \join o(s) = o(s) \ple o(o(s)). \]

So, the reuse of a non-idempotent operation from the sequential data type
would violate permutation equivalence. The simplest example of the impossibility
of reusing non-idempotent operations is the counter data type: while a
sequential counter could be trivially reused as an op-based CRDT, given
commutativity, it cannot be reused for a state-based CRDT, given that
increment is not idempotent.

\topic{Grow only set}

The classic example of a trivial state-based CRDT is the grow-only set (GSet),
shown in Figure~\ref{fig:state-gset}:
a set having only the add update, where all operations, whether
mutator (add) or queries (elements and contains) correspond to the original
sequential data type operations, and the state is simply a set, as for the
sequential data type. Merging replica states is simply the set union.
This is possible because a set is a lattice and the only update operation
(add) is an idempotent inflation.  This is an example of an \emph{anonymous
CRDT}, where there is no need for node identifiers in the state.

\begin{figure}
\begin{eqnarray*}
    \af{GSet}\typeparam{E} & = & \pow{E}\\
\bot & = & \{\} \\
  \af{add}_i(e,s) & = & s \union \{e\} \\
  \af{elements}(s) & = & s\\
  \af{contains}(e, s) & = & e \in s\\
s \join s' & = & s \cup s'
\end{eqnarray*}
\caption{State-based grow only set GSet\typeparam{E}}
\label{fig:state-gset}
\end{figure}

\topic{Single-writer principle and named CRDTs}

State-based CRDTs are less prone to have trivial implementations, namely under
the presence of non-idempotent operations. This precludes even seemingly trivial
data types, such as counters, from having correspondingly trivial state-based CRDTs.

A powerful strategy in concurrent programming is the single writer principle.
It amounts to each variable being only updated by a single process,
has long been used to obtain elegant designs, such as the Bakery
algorithm~\cite{DBLP:journals/cacm/Lamport74a}, and is specially relevant to
the design of state-based CRDTs~\cite{DBLP:conf/ecoop/EnesAB17}:
  \begin{itemize}
    \item the state is partitioned in several parts;
    \item each replica updates a part dedicated exclusively to itself;
    \item the state is joined by joining respective parts.
  \end{itemize}

Unique replica identifiers can be used to partition the state, which becomes a
map from ids to parts.  CRDTs that use replica ids in the state can be called
\emph{named CRDTs}.


\topic{State-based GCounter}

Contrary to op-based counters, state-based counters cannot use the trivial
sequential implementation, due to the non-idempotency of the increment
operation. The state-based GCounter, shown in Figure~\ref{fig:state-gcounter},
makes use of the single writer principle.
The state maps replica identifiers to integers, the $\inc$ mutator for replica
$i$ increments the self entry ($i$), and join is the pointwise max.
The counter is thus similar in structure to a version vector~\cite{DBLP:journals/tse/ParkerPRSWWCEKK83}..

\begin{figure}
\begin{eqnarray*}
  \sf{GCounter} & = & \ids \pfunc \nat \\
  \bot & = & \{ \} \\
  \inc_i(m) & = &  m\{i \mapsto m[i]+1\}\\
  \af{value}(m) & = & \sum_{j \in \ids} m[j] \\
  m \join m' & = & \{ j \mapsto \max(m[j],m'[j]) | j \in \dom m \union \dom m'
  \}
\end{eqnarray*}
  \caption{State-based GCounter}
\label{fig:state-gcounter}
\end{figure}

\topic{State-based PNCounter}

A positive-negative counter (PNCounter), having both increment and decrement
operations cannot have the same state as the GCounter, because decrement is
not an inflation. This is solved through a pair of GCounters (product
composition): increments and decrements are tracked separately. The counter
value is obtained as the difference.
This CRDT is presented in Figure~\ref{fig:state-pncounter}.
(In practice, implementations use a single map to pairs.)

\begin{figure}
\begin{eqnarray*}
  \af{PNCounter} & = & \af{GCounter} \times \af{GCounter}\\
  \bot & = & (\bot, \bot) \\
  \inc_i((p,n)) & = & (\inc_i(p), n) \\
  \dec_i((p,n)) & = & (p, \inc_i(n)) \\
  \af{value}((p,n)) & = & \af{value}(p) - \af{value}(n)\\
  (p,n) \join (p',n') & = & ( p \join p', n \join n' )
\end{eqnarray*}
  \caption{State-based PNCounter}
\label{fig:state-pncounter}
\end{figure}

\subsection{Causal CRDTs}

\topic{The problem of forgetting information}

Many times we want to remove things. An example is a set with add and
remove operations. But as we must use inflations for state mutators, in
this set example, we cannot use a single set for the state and remove
elements from it in the remove mutator. So, while a grow-only set is
trivial, this more general set is not.

A common approach to solve this limitation is to use tombstones to mark
removal, but it has the significant drawback of making the state grow forever.
An example of such approach is the 2P-Set~\cite{shapiro:inria-00555588}
(two-phase set) CRDT: it allows add and
remove, but once removed, an element cannot be re-added.
The implementation uses a pair of sets, to track adds and removes, each set
only growing, making the state always growing and eventually large after some
time, even when the number of elements considered to be in the set is small.

\topic{Causal CRDTs}

A general approach to allow removing information while avoiding tombstones
is used in \emph{Causal CRDTs}~\cite{DBLP:journals/jpdc/AlmeidaSB18}, drawing
inspiration from \emph{Dotted Version
Vectors}~\cite{DBLP:journals/corr/abs-1011-5808}.
The state has two components: a \emph{dot store} and a \emph{causal context}.
The dot store (DS) is a container for data type specific information; each
information item is tagged with a unique event identifier, a \emph{dot}, which
is a pair (replica-identifier $\times$ counter).
The causal context (CC) represents the causal history: the ids of all visible
updates, normally encoded by a version vector.

Remark: a dot is not merely a unique identifier. Being unique would be
enough for the op-based ORSet CRDT, and could have different forms, such as a
pair (Lamport-timestamp $\times$ replica-identifier), or a UUID. A dot is a
specific form of unique identifier, appropriate to be tested as
belonging to a compressed causal history in the form of a version vector.

Causal CRDTs have a pair of components for state but they are not independent,
as in a Cartesian product.  The two components are related, conveying
the knowledge that an information item with identifier covered by the causal
context and not present in the dot store has already been removed.

\topic{Joining causal CRDTs}

\begin{figure}
\begin{center}
  \includegraphics[scale=0.5]{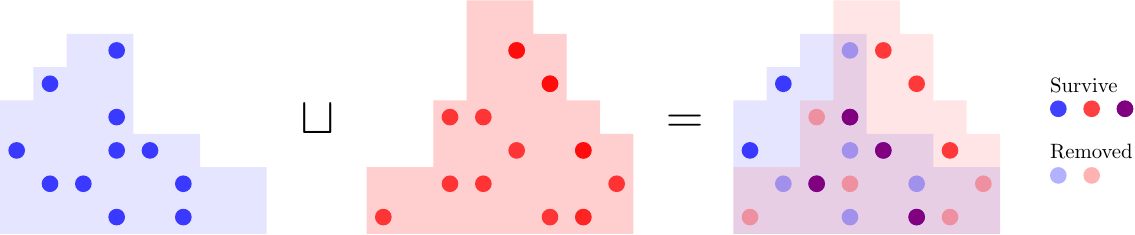}
\end{center}
  \caption{Joining causal CRDTs.}
  \label{fig:joining-causal}
\end{figure}

To merge the information represented by two replica states, the join cannot be
the standard one for product composition, i.e., we do not have the lattice
resulting from the Cartesian product of the DS and CC components, but a more
interesting one.
The join is illustrated in Figure~\ref{fig:joining-causal}.
The dot store that results from a join of $x$ and $y$:
      \begin{itemize}
        \item has dots from $x$'s DS not covered by $y$'s CC;
        \item has dots from $y$'s DS not covered by $x$'s CC;
        \item has the dots present in both DSs.
      \end{itemize}
The causal context upon a join is the join of the causal contexts.

\newcommand\auxfunpar[1]{\expandafter\newcommand\csname #1\endcsname[1]{%
  \mathop{\hbox{$\mathsf{#1}$}}\nolimits\typeparam{##1}}}
\newcommand\afpar[2]{\af{#1}\typeparam{#2}}
\auxfun{CausalContext}
\auxfun{DS}
\auxfun{DotSet}
\auxfun{Lattice}
\auxfunpar{DotMap}
\auxfunpar{DotFun}
\auxfunpar{Causal}

\topic{Some dot stores}

Being a dot store a container, we can have different types of such containers.
Three important dot stores are: $\DotSet$ -- a set of dots; $\DotFun{V}$ -- a map
from dots to values in some lattice $V$; and $\DotMap{K, V}$ -- a map from
keys in some arbitrary set $K$ to a dot store $V$:

  \begin{eqnarray*}
  \DotSet : \DS &=& \pow{\ids \times \nat} \\
  \DotFun{V : \Lattice} : \DS &=& \ids \times \nat \pfunc V \\
  \DotMap{K, \; V : \DS} : \DS &=& K \pfunc V \\
\end{eqnarray*}

DotMaps are particularly powerful, as they allow map CRDTs which embed CRDTs
as values:
  \[
    \DotMap{K_1, \; \DotMap{K_2, \; \DotFun{\pow{E}}}}
  \]

\topic{Joining causal CRDTs -- for different dot stores}

We can now define the join operator for causal CRDTs, considering the three
kinds of dot stores above. The join, shown in
Figure~\ref{fig:joining-causal-dot-stores}, simply returns the dot store
containing the appropriate surviving items, as illustrated in
Figure~\ref{fig:joining-causal}, paired with the join of the causal contexts.

\begin{figure}
  \setlength\arraycolsep{2pt}
  \begin{eqnarray*}
  \Causal{T : \DS} &=& T \times \CausalContext \\
  \join &:& \Causal{T} \times \Causal{T} \to \Causal{T} \\
  \\
  \kw{when} && T : \DotSet \\
  (s,c) \join (s',c') &=& ((s \cap s') \union (s \setminus c') \union (s' \setminus c), c \union c') \\
  \\
  \kw{when} && T : \DotFun{\_} \\
  (m,c) \join (m',c')
    &=& (\{ d \mapsto m[d] \join m'[d] | d \in \dom m \cap \dom m' \} \union \\
    && \hphantom( \{ (d,v) \in m | d \not\in c' \} \union
                  \{ (d,v) \in m' | d \not\in c \} , c \union c') \\
  \\
  \kw{when} && T : \DotMap{\_,\_} \\
  (m,c) \join (m',c') &=& (\{ k \mapsto \af{v}(k) | k \in \dom m \union \dom
m' \land \af{v}(k) \neq \bot \}, c \union c')\\
                      && \kw{where} \af{v}(k) = \fst{((m[k],c) \join (m'[k],c'))}
\end{eqnarray*}
  \caption{Join operator for causal CRDTs, considering each kind of dot store.}
  \label{fig:joining-causal-dot-stores}
\end{figure}


\topic{Observed-remove set ORSet\typeparam{E}}

To illustrate the power of causal CRDTs, we give the example of a state-based
observed-remove set, shown in Figure~\ref{fig:state-orset}, where the state
uses a DotMap from elements to DotSets.
The add mutator replaces, for the key corresponding to the element being
added, any set of dots by a new singleton.
Remove simply removes the key from the map (domain subtraction), with no need 
for the introduction of a new event in the causal context.
The causal context is in the form of a version vector.
The join, not shown, is the previously defined for the DotMap (and DotSet) case:
  \begin{itemize}
    \item it keeps concurrently added elements;
    \item it removes elements removed elsewhere if no dot survives.
  \end{itemize}

\begin{figure}
\begin{eqnarray*}
  \afpar{ORSet}{E} &=& \Causal{\DotMap{E, \DotSet}} \\
  \af{add}_i(e,(m,c)) &=& (m\{e \mapsto \{(i, c[i]+1)\}\}, c\{i \mapsto c[i] + 1\}) \\
  \af{remove}_i(e, (m,c)) &=& (\{e\} \domainsub m, c) \\
  \af{elements}((m,c)) & = & \dom m
\end{eqnarray*}
  \caption{State-based observed-remove set $\afpar{ORSet}{E}$.}
  \label{fig:state-orset}
\end{figure}

Remark: although not needed (and not done here), it would be advantageous to
generate a new event, and advance the causal context, in a remove operation,
even if there is no new dot being stored. The cost would be minuscule, for a
version vector representation of the causal context, and it would allow
trivially comparing replica versions by only comparing the causal context,
without the need to traverse the dot store, as in this implementation.

\section{Delta State CRDTs}

\topic{Operation- vs State-based approaches}

Operation-based approaches have small messages, but have strong requirements
from the underlying messaging layer: it does not work if messages are lost or
duplicated. On the other hand, state-based approaches work under weak messaging
assumptions, but send full states, which may have considerable sizes, causing
a large messaging cost.  Can we have the best of both worlds?

\topic{Delta State CRDTs}

In delta state CRDTs~\cite{DBLP:journals/jpdc/AlmeidaSB18} the state is, like
in normal state-based CRDTs, a join-semilattice, but messages are built from
\emph{delta-states}, which are states, typically small, which will make
messages themselves hopefully small.

The essential difference is that a delta CRDT has \emph{delta-mutators}, which
return delta-states, to be: 1) joined with the current state; 2) joined with other
delta-states, forming \emph{delta-groups} $d$, to send in messages.
For each mutator $m$ of a state-based CRDT we can define a delta-mutator
$m^\delta$ such that:
\[ m(X) = X \join m^\delta(X) \]

Figure~\ref{fig:state-vs-delta} summarizes the approaches.
While for normal state-based CRDTs mutators return the next state (and need to
be defined as inflations), for delta-state CRDTs delta-mutators return a value
(delta) to be joined to the current state (causing an inflation). Also,
state-based CRDTs join full states received as messages, while delta-state
CRDTs join delta-groups (join of several deltas).

\begin{figure}
  \begin{boxes}{2}
    \begin{tcolorbox}[title=State-based]
      \centering $\begin{aligned}
      X_i &\gets m(X_i) \\
      X_i &\gets X_i \join X_j \\
      \end{aligned}$
    \end{tcolorbox}
    \begin{tcolorbox}[title=Delta-state based]
      \centering $\begin{aligned}
      X_i &\gets X_i \join m^\delta(X_i) \\
      X_i &\gets X_i \join d \\
      \end{aligned}$
    \end{tcolorbox}
  \end{boxes}
  \caption{Normal state-based vs delta-state based approaches}
  \label{fig:state-vs-delta}
\end{figure}

\topic{State vs delta-state -- examples}

For each given state-based CRDT we can define one (or more) corresponding
delta-state based CRDT. The state remains the same join-semilattice, and we
only need to define corresponding delta-mutators.
Figure~\ref{fig:delta-examples} shows two examples: a GCounter and an ORSet.
The corresponding delta-mutators returns, for the GCounter, a map with just the
self-entry for the replica, instead of the full map.
For the ORSet, it returns a pair where the dot store is a singleton or empty
map, for add and remove, respectively, and the causal context has dots
representing just the element being added/removed, plus a new dot for add.

This means that delta-states will be typically very small, with either
singletons or very small sets. It also means that, unless an anti-entropy
mechanism ensuring causal-consistency is used, causal contexts in replica
states will not be downward closed and, therefore, not representable by a
version vector. (In this example causal contexts are a set of dots.)
They can, nevertheless, be represented in a compact way by a version vector
plus some extra dots.

\begin{figure}
  \newcommand\fsize{\small}
  \begin{boxes}{2}
    \begin{tcolorbox}[title=State-based GCounter]
      \fsize
     \begin{center}$\begin{aligned}
  \inc_i(m) =  m\{i \mapsto m[i]+1\}
     \end{aligned}$\end{center}
    \end{tcolorbox}
    \begin{tcolorbox}[title=Delta-state based GCounter]
      \fsize
     \begin{center}$\begin{aligned}
  \inc_i^\delta(m) = \{i \mapsto m[i]+1\}
     \end{aligned}$\end{center}
    \end{tcolorbox}
    \begin{tcolorbox}[title=State-based ORSet\typeparam{E}]
      \fsize
     \begin{center}$\begin{aligned}
    \af{add}_i(e,(m,c)) &= (m\{e \mapsto d\}, c \union d) \\
       \kw{where} d &= \{(i, \max_i(c) + 1)\} \\
    \af{remove}_i(e, (m,c)) &= (\{e\} \domainsub m, c) \\
     \end{aligned}$\end{center}
   \end{tcolorbox}
    \begin{tcolorbox}[title=Delta-state based ORSet\typeparam{E}]
      \fsize
     \begin{center}$\begin{aligned}
       \af{add}_i^\delta(e,(m,c)) &= (\{e \mapsto d\}, d \union m[e]) \\
       \kw{where} d &= \{(i, \max_i(c) + 1)\} \\
       \af{remove}_i^\delta(e, (m,c)) &= (\{\}, m[e]) \\
     \end{aligned}$\end{center}
    \end{tcolorbox}
  \end{boxes}
  \caption{State- vs delta-state examples: GCounter and ORSet.}
  \label{fig:delta-examples}
\end{figure}

\topic{Naive delta propagation algorithm}

As the standard state-based approach, delta-state CRDTs aim for allowing
arbitrary topologies.
Therefore, a basic propagation mechanism aims for transitive propagation of
information. Such an algorithm is presented in Figure~\ref{fig:basic-anti-entropy}.
It simply keeps a single \emph{delta-buffer} to broadcast to neighbors; upon some
update joins the delta produced by the delta-mutator to both CRDT state and
delta-buffer; a delta-group received in a message is joined to both CRDT state
and delta-buffer; periodically, it broadcasts the delta-buffer to neighbors
and resets it to bottom. This algorithm would even ensure causal consistency
under reliable FIFO messaging, but it fails to ensure it under the 
weaker messaging guarantees desired for state-based CRDTs.
Unfortunately, this algorithm is too naive: the delta-buffer,
even being periodically reset, easily tends to grow and become the full state.

\begin{figure}
\auxfun{send}
\auxfun{receive}
\auxfun{ack}
\auxfun{random}
  \begin{algo}{2}
\dstate{
  $X_i \in S$, CRDT state; initially $X_i = \bot$
}
\vstate{
  $D_i \in S$, delta-buffer; initially $D_i = \bot$
}
\on({$\operation_i(m^\delta)$}){
  \kwlet $d = m^\delta(X_i)$ \;
  $X_i \gets X_i \join d$ \;
  $D_i \gets D_i \join d$
}
\on({$\receive_{j,i}(d)$}){
  $X_i \gets X_i \join d$ \;
  $D_i \gets D_i \join d$
}
\periodically(){
  \kwlet $m = \af{choose}_i(X_i, D_i)$ \;
    \For{$j$ \kwin $\af{neighbors}_i$}{
    $\send_{i,j}(m)$
  }
  $D_i \gets \bot$
}
\end{algo}
  \caption{Naive propagation algorithm for delta CRDTs, for replica $i$.}
  \label{fig:basic-anti-entropy}
\end{figure}

\topic{Problems with naive delta propagation}

Unfortunately, the naive transitive propagation causes much redundancy, which
makes messages and buffers converge to the full state. Resetting the buffer
does not help if messages become large and get joined subsequently.

There are two main problems~\cite{DBLP:conf/icde/EnesAB019}: 1) deltas are
re-propagated back to where they came from; 2) a delta-group received is
wholly joined to the local delta-buffer, even if it is already mostly
reflected in the state.
The solution is a more sophisticated algorithm which: 1) avoids
back-propagation of delta-groups, by tagging the origin of each message; 2)
removes redundant state in received delta-groups (already reflected in the
local state). But how can this redundant state be identified?

\topic{Optimal deltas and smart propagation through join decompositions}

The solution to the problem of how to extract ``new information'' from a
received delta-group and also to the fundamental problem of how to define
optimal delta-mutators (that produce the smallest delta possible) can be found
in results in lattice theory developed by~\textcite{birkhoff1937},
namely the concept of \emph{join decompositions}.

The main concepts and results are summarized in Figure~\ref{fig:join-decompositions}.
An element is said to be join-irreducible if it cannot result from the join
of a finite number of elements not containing itself. A join decomposition of
some element $x$ is a set of join-irreducible elements whose join produces
$x$. An irredundant join decomposition (IJD) is when no element is redundant
(removing any element produces a smaller result when joined).

\begin{figure}
\newcommand\decomp[1]{{{\Downarrow}#1}} 
  \begin{boxes}{2}
    \begin{tcolorbox}[title=Join-irreducible]
    $ x = \bigjoin F \implies x \in F $
    \end{tcolorbox}
    \begin{tcolorbox}[title=Join decomposition]
    $ D \subseteq \mathcal{J}(L) \land \bigjoin D = x$
    \end{tcolorbox}
    \begin{tcolorbox}[title=Irredundant join decomposition (IJD)]
    $ D' \subset D \implies \bigjoin D' \ple \bigjoin D $
    \end{tcolorbox}
    \begin{tcolorbox}[title=Unique IJD for distributive lattices]
    $ \decomp x = \max \{ r \in \mathcal J(L) | r \pleq x \} $
    \end{tcolorbox}
    \begin{tcolorbox}[title=Difference]
    $ \Delta(a, b) = \bigjoin \{ y \in \decomp{a} | y \not \pleq b \} $
    \end{tcolorbox}
    \begin{tcolorbox}[title=Optimal delta-mutator]
    $ \af{m}^\delta(x) = \Delta(\af{m}(x), x) $
    \end{tcolorbox}
  \end{boxes}
  \caption{Join decompositions and optimal deltas.}
  \label{fig:join-decompositions}
\end{figure}

Birkhoff's representation theorem establishes a correspondence between an
element of a finite distributive lattice and the downward closed set of the
join irreducibles below it. This down set is isomorphic to the set of its
maximals, which is the unique irreducible join decomposition. (A distributive
lattice is a poset with not only the join but also its dual \emph{meet}, and
where one distributes over the other.)

For most CRDTs, the state is not merely a join-semilattice, but a distributive
lattice. Even if the state normally belongs to infinite lattices, we can apply
the result to CRDTs~\cite{DBLP:conf/icde/EnesAB019} applying it to finite
quotient sublattices. This allows obtaining the unique IJD of $x$, as the
maximals of the set of irreducibles below $x$.

Having unique IJDs, we can define the difference between two states $a$ and $b$:
it is the join of the elements in the unique IJD of $a$ not below $b$.
(As an example, the difference becomes simply set difference when the lattice is a
powerset.)
The difference to the current state can be used to extract ``new'' information
from a received delta-group, and the optimal delta-mutator can be defined as
the difference between what the original mutator produces, $m(x)$, and the
current state $x$.

\topic{Inter-datacenter delta propagation in geo-replicated deployments}

Delta propagation algorithms for the general case of a network of nodes
require considerable care to be efficient.
But practical deployments of CRDTs may be in specific more restricted
scenarios, that can be exploited to simplify communication. One
such case is having just a few replicas, one in each of a small number of
nodes (e.g., datacenters) connected in a full mesh.

In this case, the system can run, by default, without transitive
propagation: each replica will only send its own deltas to all others.
A few simple ingredients will allow an efficient causal delta-propagation algorithm:
merging only the self-issued deltas in delta-groups; starting a new delta-group after
applying a received remote one; keeping track (with a version vector) of the
logical time when each delta-group starts, to send it together with the
delta-group; and delay applying received delta-groups until causal dependencies
are met.
This avoids the need to perform join-decompositions to achieve
efficiency, as in a more general delta-propagation algorithm. 
It amounts to performing a kind of causal broadcast, where the unit of
propagation is a delta-group, with the advantage of 1) allowing compression
due to delta-merging; 2) the possibility of, given a long network partition,
discarding the delta-groups and falling back to sending the full-state when
the partition heals.

\section{Comparison of the approaches}

A comparison of the approaches is now made, regarding issues such as
open vs closed systems, partition tolerance, state size (including the issue
of amortizing metadata overhead over several objects). A summary of these
aspects (as well as other already discussed) is presented in
Table~\ref{tab:comparison}. We start by a taxonomy, discussing the
dual nature of some CRDTs.

\topic{Taxonomy of CRDTs}

Some CRDT cannot be state-based, as the state is not a join-semilattice, not
supporting merging states, or effect is not an inflation.
On the other hand, any state-based CRDT could be considered, vacuously, op-based, by
defining $\prepare$ as the state mutator and $\effect$ as the join operation.
But this is not useful for taxonomy purposes and would make no sense in
practice. Therefore, we should classify as op-based CRDTs only those in which
$\prepare$ does not return the full state.
Interestingly, delta-state CRDT implementations can be repurposed to
the op-based model, fitting this restriction, and can be considered to have a
dual nature as both state- and op-based. Given operation $o$, corresponding
delta mutator $m_o^\delta$, and current state $s$ we can define
  $\prepare(o, s) \defeq m_o^\delta(s)$ and $\effect \defeq \join$.
This means that we can use op-based propagation for delta-state CRDTs. It
would be, however, typically worse, as no delta-merging before propagation
could be exploited, and delta-CRDTs do not need the stronger delivery
guarantees used for op-based propagation.

Conversely, for some op-based CRDTs the state happens to be a
join-semilattice, $\prepare$ returns an element of that lattice, and $\effect$
is the same for all operations and amounts to the lattice join. Such CRDTs,
even if originally defined under the op-based model, can switch to use a
delta-based propagation model (possibly requiring causal delta-propagation,
depending on the case), gaining flexibility and possibly higher efficiency,
defining
  $m_o^\delta(s) \defeq \prepare(o,s)$ and $\join \defeq \effect$.
Going in this direction is practically useful, not just a taxonomic
curiosity. Some List CRDTs, discussed in Section~\ref{sec:colab-editing}, have
this dual nature and can choose either model.

Pure op-based CRDTs are a special case. In the naive form, without \polog
compaction, have a lattice for state (set of causally tagged operations) and
the result of prepare, tagged with causality information, can be considered a delta.
We can define
  $m_o^\delta(s) \defeq  \{(t, o)\}$ and $\join \defeq \union$,
where $t$ is the causal timestamp assigned to operation $o$. For this we need
to expose the causal context present in the TCB middleware to the CRDT, to
generate timestamps at the source.
This means that naive pure-op based CRDTs can use a more flexible delta-state
based propagation.  They would typically need causal delta propagation, for
semantic reasons and also to make causal tagging simple.
\polog compaction can also be achieved in this model, but requires some
care. Namely, we cannot simply use the set of tagged operations for lattice as
compaction would cause a deflation. This would cause removed operations to
reappear, incorrectly, upon a merge. But we can make the set of tagged
operations a Dot Store and, together with the exposed causal context obtain a
causal CRDT, where compactions become inflations, and merge works correctly.

\begin{figure}
\begin{center}
  \includegraphics[scale=0.6]{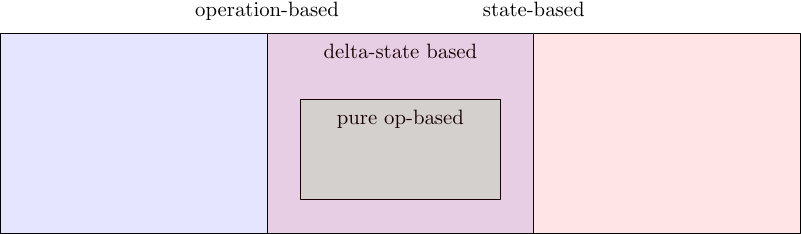}
\end{center}
  \caption{Taxonomy of CRDTs, showing the dual nature of some CRDTs as both
  op-based and delta-state based, including pure op-based CRDTs.}
  \label{fig:CRDT-taxonomy}
\end{figure}

Given these observations, we obtain the taxonomy of CRDTs presented in
Figure~\ref{fig:CRDT-taxonomy}.

\begin{table}
  \caption{Comparison between CRDT approaches.}
  \label{tab:comparison}
  \begin{center}
    \small
    \begin{tabular}{@{}lcccc@{}}
  \toprule
      & \multicolumn2c{operation-based} & \multicolumn2c{state-based} \\
   \cmidrule(l){2-3} \cmidrule(l){4-5}
      & general & pure-operation & standard & delta-state \\
  \midrule
  reuse sequential ADTs &
    \multicolumn2c{commutative ops} & \multicolumn2c{idempotent inflations} \\
  open vs closed systems &
      \multicolumn2c{fixed set of participants} & \multicolumn2c{dynamic
      independent groups} \\
  partition tolerance &
      \multicolumn2c{unsuitable for long partitions} & \multicolumn2c{good} \\
  state size &
      \multicolumn2c{independent of replicas} & 
      \multicolumn2c{component linear with replicas} \\
  metadata amortization &
      \multicolumn2c{transparent (middleware)} & \multicolumn2c{explicit (container)}  \\
  message size & normally small & small & large & possibly small \\
  messaging (usual) & \multicolumn2c{causal broadcast} & 
      \multicolumn2c{epidemic at-least-once} \\
  causal consistency & \multicolumn2c{from messaging} & for free & needs care \\
      causal stability & useful & important & \multicolumn2c{not available
      in general} \\
  \bottomrule
    \end{tabular}
  \end{center}
\end{table}

\topic{Open vs closed systems}

Operation-based CRDTs, using causal broadcast, assume a known set of
participants. Dynamic joining or retiring of replicas can possibly be done but it is a
complex operation, needing some care. But two independent groups of replicas
that have evolved independently cannot merge. (For realistic
implementations, not considering naive implementations that keep all history
of operations, or without becoming effectively state-based CRDTs.)
On the other hand, state-based CRDTs suit an unknown set of participants, and
allow trivial dynamic joining, leaving or merging of independent groups of
replicas. The only requisite, for named CRDTs, is to have globally unique
replica identifiers, a reasonable assumption.

\topic{Partition tolerance}

Both kinds of CRDTs are network partition tolerant in terms of operations
being always available, locally, but a partition has significantly different
consequences for each case.
In operation-based CRDTs there is a transient cost which is linear with the
partition duration, having to do with the need to store operations not yet
delivered to all. Thus, they may be unsuitable for large weakly connected
systems prone to long partitions, which will cause unbounded memory
consumption.
State-based CRDTs are better in this regard, as each replica is completely
autonomous, with no need to track individual operations. Each operation is
immediately incorporated in the state, and the approach works well even under
long network partitions.
State-based CRDTs are thus more suitable for autonomous operation over
unstructured networks with poor connectivity.

\topic{State size}

Operation-based CRDTs can frequently have smaller states, independent of
the number of replicas, even having non-idempotent operations (e.g., a counter). 
This comes from the power from delegating to a reliable message delivery 
mechanism. Even though the size can be in theory, in the worse case, linear
with the number of replicas, in practice that is not an issue. E.g., an ORSet with
replicas issuing concurrent adds on the same elements could have size
temporarily linear with the degree of replication, but in practice that would
be unlikely to happen for a substantial proportion of elements in the set at
the same time; moreover, subsequent operations would reduce state size,
bringing it to an effectively constant small size per element.

State-based CRDTs assume the worst case of unreliable messaging. This leads
normally to some state component that grows linearly with the number of
replicas. That may be very relevant, e.g., for a counter, where having a map
from replica ids to integers is much more space than a single integer; or less
relevant, e.g., for a large set, implemented as a causal CRDT with a
DotMap, where the cost of the causal context pales in comparison with the
large number of elements in the set and where the meta-data cost per set
element (a set of dots, frequently a singleton) is small.
In general, we can say that for small data types (scalar-like, e.g., counters,
or small container-like, e.g., sets), state-based CRDTs have significant
overhead (linear with the number of replicas) while for large container-like
data types (sets, maps, lists) the overhead for state-based CRDTs becomes less
of a problem.

\topic{Amortizing metadata overhead over many objects}

As several objects of a data type are frequently used together, another
question is whether the overhead in meta-data can be amortized over several
objects. In this respect there is some difference in the approaches.
In operation-based CRDTs, the cost (e.g., a version vector in some
implementations of causal broadcast) is independent of the number of CRDT
objects and is amortized by all objects (even of different CRDT types) that
use the middleware. This cost can be almost none for operation-based CRDTs
that do not require knowledge about happens-before (like for counters, or that
extract relevant information from the CRDT state itself in prepare), e.g.,
using a causal broadcast based on a tree topology with FIFO
channels~\cite{DBLP:conf/eurosys/BravoRR17}, that requires negligible metadata
even for a large number of participants. Pure op-based CRDTs cannot use this
strategy, as they need knowledge about happens-before to be provided
by the causal broadcast middleware (prepare is unable to extract information
from the CRDT state).
State-based CRDTs not only have non-amortizable metadata cost per object
in some cases (like counters), but cannot have a transparent amortization as
op-based CRDTs can. For causal CRDTs (like an ORSet), it is possible to amortize
metadata cost over many objects, but only by explicitly putting those
objects inside a container.
E.g., if we have many ORSets, it is possible to place them inside a map CRDT,
thereby sharing a single causal context over those many objects. But this is
not transparent, needs a refactoring effort, being undesirable from the
software engineering perspective.

\section{Identity management towards scalability}

Apart from simple, anonymous CRDTs such as GSets, most CRDTs make use of
unique replica identifiers, with each replica using its own identifier to
generate unique ids, in the form of dots, and updating part of the state that
``belongs to it''. Causal CRDTs use a causal context in the form of a version
vector, whose size is linear with the number of replicas. State-based counters
have a similar state. This makes the growth of maps where the keys are replica
identifiers the main obstacle to scaling CRDTs to a large number of
participants. We discuss here some techniques that can be used to address it.

\subsection{CRDT-based Identity management}

\topic{Identity containment and handoff}

For CRDTs such as small ORSets or for counters, the size of the set of ids
determines the number of map entries and, therefore, the state size. Scaling a
state-based counter CRDT to many participants is a difficult problem. It was
addressed in Handoff Counters~\cite{DBLP:journals/dc/AlmeidaB19} by using a
hierarchical structure, in which only the entries of the small number of
replicas at tier 0 (e.g., datacenters) are propagated to all replicas. Any
other state generated using the id of replicas from other tiers is just
temporary, being propagated under peer-to-peer interactions but remaining
contained, not propagated to the whole system, and being eventually garbage
collected through local interactions.  The more intricate aspect is the
handoff, which migrates accounted values towards lower tiers, making all
increments eventually be accounted in the entries for tier 0 ids. This approach is
more generally applicable to other data types with associative and commutative
operations.

\topic{Identity borrowing}

Handoff Counters are considerably complex, and the generality of allowing many
tiers may not be needed. Borrow Counter~\cite{DBLP:conf/eurosys/EnesBA017} is
a simpler proposal, being a Causal CRDT that only distinguishes permanent
replicas (typically servers) and transient replicas. 
A transient replica $i$ may ask a permanent replica $j$ to create a dot,
based on the permanent replica id $j$, for $i$ to use as key in a map where to
store increments within values. The transient replica may retire by flagging
borrowed dots as inactive, upon which they are able to be acquired by
$j$ and garbage collected. The permanent state, namely the causal context,
grows only with the number of permanent replicas.

\topic{Renaming operation unique identifiers}

While the two previous approaches exploit the fungibility of increments,
causal CRDTs such as ORSets identify individual operations using dots.
For these, a possibility towards scaling, assuming a simple classification of
replicas as clients or servers, is to: 1) rename dots based on client ids to
become based on server ids; 2) only propagate globally the server-based dots.
The causal context will only depend on the number of servers, achieving
scalability to many clients.
This approach is still under research; \textcite{MarubayashiBaquero23} provide
an example of a preliminary development.

\subsection{Server-based identity management}

\topic{Identity reuse and collapsing identities}

Another possibility is to exploit system assumptions, externally to CRDTs themselves.
An example: clients obtain replicas from a central server/service to use in a
session involving peer-to-peer interactions; a client uses the replica during
the session and then relinquishes the right to use it, handing it
back to the server. A strongly consistent service can control replica
use by clients, assign ids to clients, keep track of ids in use and the ones
that became inactive (when clients end sessions), and only assign new ids when no
previously generated id is inactive.
This allows the size of the id set to grow only with the number of
clients that concurrently interact, and not the number of clients that have
ever used the CRDT. This keeps the benefits of CRDTs, namely allowing local
peer-to-peer interactions. Partitions may only delay the possibility of
reusing an identity, which in the worst case will cause a new identity to be
generated. Moreover, if no session is active, for most state-based CRDTs that
we discussed up to now, the server could collapse all identities and dots into
a single one. This would require an extra $\af{collapse}$ operation to be
provided in the CRDTs which, e.g., for a counter would sum all the entries and
return a map with a single entry. We are not aware of these techniques having
been described or if they have been used.

\section{Practical applications}
\label{sec:practical-applications}

CRDTs started being used in the industry for scenarios involving large scale
or partition tolerance in a distributed setting.
Early examples were in the League of Legends~\cite{HS-LoL} game
by Riot Games, using Riak CRDTs and implementing an Ejabberd CRDT library, and
in SoundCloud's Roshi~\cite{Roshi}, a time-series event storage via a
LWW-element-set CRDT.
CRDTs also became supported in libraries or frameworks.
Akka is a popular framework for building distributed applications,
with emphasis on supporting the actor model, having added
support~\cite{Akka-CRDTs} for state based CRDTs, both standard and
delta-state.
An area that motivated much research and resulted in successful products
was NoSQL databases.
Pre-CRDT NoSQL data stores, such as Amazon's Dynamo, were difficult to use due
to the need for ad hoc reconciliation of concurrent writes.
Another area of application where CRDTs were successful has been collaborative
editing. We now present examples in these two areas.


\subsection{Databases}

Following Amazon's Dynamo, Riak~\cite{10.1145/1900160.1900176} was one of the
first successful data stores which added support for CRDTs through Riak DT,
which introduced several CRDTs, such as the Riak DT
Map~\cite{DBLP:conf/eurosys/BrownCME14}, possibly the first map
where values themselves could be state-based CRDTs. It also provided the
ORSet. Other applications were designed on top of Riak Core, such as Riak
PG~\cite{DBLP:conf/erlang/Meiklejohn13}, a process group registry for Erlang using ORSets.
Riak DT ORSets were used successfully by bet365~\cite{bet365}, an online
gambling site, to scale their application.
Redis is a very popular in-memory database with an emphasis in being a data
structure store. It added support for active-active geo-distribution and
CRDT-enabled databases~\cite{Redis-CRDTs}.
AntidoteDB~\cite{AntidoteDB} is a database which served as reference platform and
applies research from two European Projects, SyncFree and LightKone.
It provides highly-available transactions and geo-replication together with CRDTs.
A related database, SwiftCloud~\cite{DBLP:conf/srds/PreguicaZBDBBS14}, also
supports transactions over CRDTs, while supporting client-side partial
replication and local execution. These systems allow transactions that read
from causally consistent snapshots and apply updates atomically, exploiting
CRDTs to allow merging updates from different transactions. This allows
committing transactions that would have to abort under strong consistency requirements.

\subsection{Collaborative editing -- the List data type}
\label{sec:colab-editing}

Collaborative editing is possibly one of the best examples of a topic
involving a non-trivial data structure (List), where CRDTs achieved
considerable success. The long line of research before CRDTs faced many difficulties and
failed to overcome problems for the non-centralized scenario in a satisfactory way.
It is well worth contrasting pre-CRDT and CRDT solutions for the
List~\cite{DBLP:conf/podc/AttiyaBGMYZ16} data type. A possible API contains
two update operations:
$\ins(e, k)$ inserts an element $e$ at position $k$ (index, starting from 0), and
$\del(k)$ deletes element at position $k$.

\topic{Before CRDTs -- Position Transformation}

For the List data type, given two operations concurrently issued at
different replicas, applying them in different serial orders in different
replicas does not lead to convergence nor the intended result. The first
approach to address this problem was to transform an operation before applying
it remotely, so as to obtain convergence and also the intended
effect as when it was issued at the originating replica.
This approach became known as the OT (Operational Transformation)
approach, from the \emph{Distributed Operational Transformation
(dOPT)}~\cite{DBLP:conf/sigmod/EllisG89} algorithm.
Transformations modify positions (indexes) of operations; e.g., an
$\ins(x, 5)$ issued at some replica $i$ could be transformed to, say, $\ins(x,
7)$ before being applied at another replica $j$, to maintain the intended
effect, given concurrent modifications at $j$ that inserted two elements before position 5.

Unfortunately, transforming operation positions appropriately turned out to be
much more difficult, complex and error-prone than expected. While it is
relatively manageable under centralized
coordination~\cite{DBLP:conf/uist/NicholsCDL95}, e.g., in Google
Docs, it becomes dauntingly complex in the general case of non-centralized peer-to-peer
replica synchronization, which is important for \emph{local-first}
applications~\cite{DBLP:conf/oopsla/KleppmannWHM19}.

Problems with the original dOPT algorithm were identified
by~\textcite{DBLP:conf/cscw/ResselNG96}. They proposed a framework involving
two \emph{transformation properties} ($TP_1$ and $TP_2$), to be satisfied by
\emph{transformation functions} so as to ensure correctness. $TP_2$ turned out
very difficult to satisfy and~\textcite{DBLP:conf/ecscw/ImineMOR03} showed
that all initially proposed transformation functions did not satisfy it, and
proposed different transformation functions. Unfortunately, these were found
to violate $TP_2$ and intention preservation~\cite{DBLP:conf/cscw/LiL04}.
Another proposal~\cite{DBLP:conf/colcom/OsterMUI06} enforces both $TP_1$ and
$TP_2$, by relying on maintaining a \emph{tombstone} when an element is
deleted, defining \emph{Tombstone Transformation Functions}.

\begin{exclude}
An approach somewhat dual to OT, instead of transforming operations towards
being applied to a document, transforms the document back to the operation.
The \emph{Mark \& Retrace}~\cite{DBLP:conf/group/GuYZ05} algorithm implements an \emph{AST
(Address Space Transformation)}: to apply an operation to some replica, this
technique transforms (retraces) the document address space to reflect the
state back at the time when the operation was issued.


In addition to the extreme difficulty of transforming operation positions,
these approaches suffer a major drawback: the transformation depends on the
exact causal relation between operations (i.e., whether they are concurrent or
causally related). This means that it is not enough to merely apply operations
respecting causality, for which a Lamport
clock~\cite{DBLP:journals/cacm/Lamport78} would be enough, but each element in
the document needs to be precisely tagged with a version
vector~\cite{DBLP:journals/tse/ParkerPRSWWCEKK83}, and these have a size
linear with the number of replicas, hindering scalability. (One notable
exception to this requirement is the MOT2~\cite{DBLP:conf/colcom/CartF07}
algorithm.)
\end{exclude}

\topic{CRDTs approaches -- use immutable globally unique element identifiers}

To avoid the complexity and fragility of transforming positions,
CRDT approaches assign globally unique identifiers to elements, and use them
in operations, instead of the position. These ids can be
sent in messages and used in other replicas to refer to elements, even if
their position has changed.
These ids can be an explicit part of the API (``insert after element with id
$x$''), or they can be encapsulated in the data type representation
and retrieved before being sent in messages. In a position-based API an 
``insert at position $k$'' can be implemented as ``insert after id
of element at position $k-1$'', or ``before id of element at position $k$'', or
``between the ids of elements at positions $k-1$ and $k$''.

CRDT approaches to Lists use basically a growing set of operations involving
unique ids for state (even if in practice performance optimizations are done,
such as having hash-tables and linked-lists).
They can be seen as both state-based CRDTs (state is a set) or
operation-based CRDTs (propagate operations and apply effect). As adding an
operation to a set is idempotent, the operations can also be seen as deltas,
and exactly-once delivery is not needed. But as operations may refer to ids
in other operations in their causal past, practical implementations may need
causal delivery. Nevertheless, this dual state- and op-based nature allows
these CRDTs to be used with either kind of propagation; e.g., normally
op-based propagation to have low latency of visibility, but also allowing
long partitions and offline editing, by updating local state and doing
a state merge when the partition heals.

Collaborative editing based on CRDTs has achieved notable success and is being
used in popular libraries, such as Automerge~\cite{Automerge}, based on
RGA~\cite{DBLP:journals/jpdc/RohJKL11}, and Yjs~\cite{Yjs}, based on a
modified version of YATA~\cite{DBLP:conf/group/NicolaescuJDK16}, with a list of over 40
applications using it.
The two main variants of these approaches use either some pre-existing dense
total order or build a total order as operations are issued.

\topic{CRDTs with a dense total order of globally unique identifiers}

A particular approach generates ids from a totally ordered set.
An insert becomes essentially generating an id between the ids of
the ``before'' and ``after'' elements. To make an insert always possible,
this approach uses a \emph{dense total order}, where for any two identifiers $x < y$,
there exists a $z$ such that $x < z < y$. A way to have a dense domain is
using lists for ids, representing paths in a tree. To ensure
globally unique id generation, element ids also contain replica
ids, to tie-break when two replicas insert concurrently at the same position.
Some CRDTs which use this approach are Treedoc~\cite{DBLP:conf/icdcs/PreguicaMSL09},
Logoot~\cite{DBLP:conf/icdcs/WeissUM09}, and
LSEQ~\cite{DBLP:conf/doceng/NedelecMMD13}.
\begin{exclude}
\emph{Partial Persistent Sequences}~\cite{DBLP:conf/icde/WuPF10} is a CRDT that uses
rational numbers for ids and different predefined sub-division ranges for
different replicas, to try to achieve uniqueness while avoiding replica ids.
This will not, however, work as intended (only one insert per replica at the
same position is considered, not arbitrary sequences of inserts involving many
replicas before synchronization with another).
\end{exclude}
A problem with these approaches is that the size of generated ids keeps growing over time.

\topic{CRDTs that explicitly order unique identifiers}

To avoid ever-growing element ids, some approaches do not assign ids from a
pre-existing dense total order, but: 1) use simple small globally unique ids,
such as pairs (replica id, counter), what we call a dot; 2) explicitly add
ordering constraints (rules), such as ``$(i,4)$ is before $(j,6)$''.
Ordering constrains are only added but not changed or removed. A delete
operation does not remove the element or constraints but creates a tombstone.
This allows small sizes and simple convergence, considering the state as an
ever growing set of operations and immutable constraints.
Some CRDTs that follow this approach are
Woot~\cite{DBLP:conf/cscw/OsterUMI06}, Causal
Trees~\cite{DBLP:conf/wikis/Grishchenko10},
RGA~\cite{DBLP:journals/jpdc/RohJKL11}), and
YATA~\cite{DBLP:conf/group/NicolaescuJDK16}.
A problem with these approaches is the accumulation of tombstones in the
state. Tombstone removal can be achieved resorting to causal stability. This
is in fact done algorithmically in RGA, without the authors using a term for
it, while citing~\textcite{Golding92} as already introducing a safe condition for
removal. In fact, that thesis discusses log pruning using classic message
stability (message received everywhere) and not causal stability (no more
concurrent messages delivered). This is another example of the confusion
between these concepts.

\topic{Interleaving anomalies}

The List data type also illustrates that achieving convergence maybe the
easiest part of a CRDT. Achieving, and even defining, the intended outcome given
concurrent updates can be far more problematic. Different List CRDTs can
produce strange unintended outcomes given concurrent updates. An example
from~\textcite{DBLP:conf/eurosys/KleppmannGMB19} is concurrent insertions at
the same positions: given an initial list ``\verb|Hello!|'', a user inserts
``\verb| Alice|'' before ``\verb|!|'' (6 inserts) and another user
concurrently inserts ``\verb| Charlie|'' (7 inserts), also before
``\verb|!|''. Many CRDTs can produce an outcome, after converging, such as
``\verb|Hello Al Ciharcliee!|'', exhibiting an \emph{interleaving anomaly},
when a more desirable outcome would be ``\verb|Hello Alice Charlie!|''.
Moreover, this anomaly is allowed by the specification
in~\textcite{DBLP:conf/podc/AttiyaBGMYZ16},
which~\textcite{DBLP:conf/eurosys/KleppmannGMB19} argue being too
weak for collaborative editing, and propose a stronger specification.
The anomaly manifests in Logoot, LSEQ, Treedoc and Woot; a lesser anomaly
occurs in RGA, if typing backwards.
\textcite{DBLP:journals/corr/abs-2305-00583} present a detailed study of the
interleaving problem as well as a CRDT (Fugue) that solves it.

\section{Final remarks}

We came a long way from ad hoc optimistic replication. CRDTs offer a
principled approach to optimistic replication, achieving semantically
meaningful convergence and causal consistency, two main goals for partition tolerant systems.
They avoid ``losing concurrent updates'' without requiring concurrency control
or transactions, and allow some desirable outcomes not possible if restricting
to sequential histories.
Like sequential data types, they provide general useful abstractions, but
their usage may be more difficult, as they may embrace concurrency in their
specification and force programmers to think in terms of concurrency.
Nevertheless, CRDTs have been successfully applied in
the industry, allowing systems with low response times even for large spatial spans.

The operation-based and state-based are substantially different approaches.
The former is more suitable to a known group of participants under good
connectivity, and the latter is good for long partitions, disconnected operation
and independent groups of participants that eventually meet. The op-based
approach allows small messages, better reuse of sequential data types (with
commutative operations), and tends to have smaller state. The state-based approach
results in typically larger states and messages, leading to less frequent
propagation, and less ``fresh'' data.  This is partially solved by the
delta-state approach which, if using smart delta propagation, can potentially
result in smaller messages, allowing more ``freshness''. The pure-op approach
is a special point in the design space of op-based CRDTs, demanding causality
information from the middleware.
It is limited compared with the general op-based approach, as it cannot extract
information from the state when operations are invoked. Causal stability is an
important ingredient in pure-op CRDTs, to optimize state size, but nothing
prevents it from being used in the general op-based approach, where it
may be useful, e.g., to discard tombstones, as it provides semantically
useful information.

One frequent criticism is that it is difficult to obtain correct and
efficient implementations. The answer to such criticism is that, like for
sequential data types, they need only to be implemented by a few experts,
while many practitioners can reap the benefits. This is the reason why, 70
years after being invented, research still goes on in Hash
Tables~\cite{10.1145/3625817}. If CRDTs keep revealing themselves useful,
the need for research and implementation effort will not be the problem.

\section*{Acknowledgments}

I would like to thank the anonymous reviewers for their comments which helped
improve the paper.
This work is financed by National Funds through the Portuguese funding agency, FCT - Fundação para a Ciência e a Tecnologia, within project UIDB/50014/2020.
DOI: \href{https://doi.org/10.54499/UIDB/50014/2020}{10.54499/UIDB/50014/2020}

\printbibliography

\end{document}